\documentstyle[12pt,preprint]{aastex}

\newcommand{\II}{{\scshape~ii\@}}
\newcommand{\III}{{\scshape~iii\@}}
\newcommand{\V}{{\scshape~v\@}}

\begin{document}

\title{Star Formation in a Complete Spectroscopic Survey of Galaxies}

\author{B. J. Carter, D. G. Fabricant, M. J. Geller, M. J. Kurtz}
\affil{Center for Astrophysics, 60 Garden Street, Cambridge, MA 02138}
\and
\author{B.~McLean}
\affil{Space Telescope Science Institute, Baltimore, MD 21218}

\begin{abstract}
The 15R-North galaxy redshift survey is a uniform spectroscopic survey
(S/N $\sim $10) covering the range 3650---7400\AA\ for 3149 galaxies
with median redshift 0.05.  The sample is 90\% complete to
$R=15.4$. The median slit covering fraction is 24\% of the galaxy,
apparently sufficient to minimize the effects of aperture bias on the
EW(H$\alpha$).

Forty-nine percent of the galaxes in the survey have one or more
emission lines detected at $\geq 2 \sigma$.  In agreement with
previous surveys, the fraction of absorption-line galaxies increases
steeply with galaxy luminosity.

We use H$\beta$, O[III], H$\alpha$, and [N\II] to discriminate between
star-forming galaxies and AGNs. At least 20\% of the galaxies are
star-forming, at least 17\% have AGN-like emission, and 12\% have
unclassifiable emission. The unclassified 12\% may include a
``hybrid'' population of galaxies with both star-formation and AGN
activity.  The AGN fraction increases steeply with luminosity; the
fraction of star-forming galaxies decreases.

We use the EW(H$\alpha$ + [N\II]) to estimate the Scalo birthrate
parameter, $b$, the ratio of the current star formation rate to the
time averaged star formation rate.  The median birthrate parameter is
inversely correlated with luminosity in agreement with the conclusions
based on smaller samples \citep{Kennicutt94}. Because our survey is
large, we identify 33 vigorously star-forming galaxies with $b >
3$. We confirm the conclusion of \citet{Jansen01} that EW(O[II]) must be
used with caution as a measure of current star formation.

Finally, we examine the way galaxies of different spectroscopic type
trace the large-scale galaxy distribution. As expected the absorption
line fraction decreases and the star-forming emission-line fraction
increases as the galaxy density decreases. The AGN fraction is
insensitive to the surrounding galaxy density; the unclassified
fraction declines slowly as the density increases. For the
star-forming galaxies, the EW(H$\alpha$) increases very slowly as the
galaxy number density decreases.

Whether a galaxy forms stars or not is strongly correlated with the
surrounding galaxy density averaged over a scale of a few Mpc. This
dependence reflects, in large part, the morphology-density relation.
However, for galaxies forming stars, the stellar birthrate parameter
is remarkably insensitive to the galaxy density.  This conclusion
suggests that the triggering of star formation occurs on a smaller
spatial scale.

\end{abstract}
\keywords{
   galaxies: active --- 
   galaxies: fundamental parameters (classification, colors) --- 
   galaxies: distances and redshifts ---
   galaxies: interactions ---
   galaxies: spectroscopy --- 
   galaxies: starburst
}

\section{Introduction}

Although redshift surveys have come of age, only a few recent surveys
have sufficiently uniform, high quality spectra to examine the
spectral properties of galaxies in the nearby universe \citep{Tresse,
Ho97, Bromley, Folkes, Blanton}.  The interrelations among spectral
properties, morphology, and intrinsic luminosity of galaxies are
important for understanding their formation and evolution. The spatial
distribution of galaxies as a function of spectral type is also a test
of models for galaxy and structure formation.

Some broad general conclusions on these issues are robust from one
survey to another. The fraction of emission-line galaxies increases
with decreasing intrinsic galaxy luminosity \citep{Tresse, Lina,
Zucca}.  Over a broad range of luminosities, emission-line galaxies
appear less strongly clustered than absorption-line galaxies
\citep{Loveday99, Iovino, Salzer, Rosenberg, Linb}.

Aperture bias may be important in assessing the relative distributions
of star-forming and absorption-line galaxies \citep{Kochanek}.
Apertures which cover only a small fraction of a galaxy, such as in 
fiber-fed spectroscopy, may
bias the classification against emission line objects by missing HII
regions in spiral disks or in the outer regions of irregulars. The
larger the bulge of the galaxy, the more potentially serious the bias. The
well-known morphology density relation along with a small aperture can
conspire to produce a stronger dependence of the star-formation rate
on density than is actually present.

A detailed understanding of the emission-line population at zero redshift
is an important cornerstone for assessing the star formation history
of the universe. A number of investigators have concluded that the
star formation rate has decreased substantially from z $\sim$ 1.5 to
the present. One uncertain aspect of the star fromation rate at zero
redshift is the possible large contribution of AGNs to the
emission-line population.

The quality of the spectra and the inclusion of the H$\beta$, [O\III],
H$\alpha$, and [N\II] lines are important for discriminating between
AGN and star-forming galaxies.  \cite{Ho97} find AGNs in 44\% of their
sample of 471 high quality (S/N $\sim$ 100) nuclear spectra of nearby
galaxies which include these diagnostic lines.  Surveys with poorer
spectra and/or more limited spectral coverage reveal AGNs at much
lower rates \citep{Huchra, Bothun, Tresse, LCRS}.  The Las Campanas
Redshift Survey \citep{LCRS} spectra do not include H$\alpha$ and [N\II]; the
Stromlo-APM \citep{Tresse} spectra do not include [O\III]. If the AGN
fraction is truly large, unidentified AGNs can bias the assessment of
the star formation rate.

To examine the properties and distribution of star-forming galaxies in
the nearby universe, we discuss the 15R survey, a uniform,
complete spectroscopic survey of 3149 galaxies with $R \leq 15.4$.
Our spectra cover the entire wavelength range 3650---7400\AA\
with typical S/N $\sim$ 10 and include H$\alpha$ and [N\II]. The median
fraction of the integrated galaxy light included in our spectra is 24\%.

Section 2 describes the details of the survey. We demonstrate that the
equivalent width (EW) of H$\alpha$ is insensitive to aperture bias.
Section 3 discusses
the application of the Veilleux-Osterbrock (1987) classification
scheme to discriminate between star-forming galaxies and those
dominated by an AGN{}.  We use the EW(H$\alpha$ + [N\II]) to estimate
the Scalo birthrate parameter, a measure of the current star-formation
rate. We identify a set of 33 galaxies which are undergoing an intense
burst of star formation. We also explore the use of [O\II] as a
indicator of star formation.  In section 4 we explore the relationship
between local density and the spectral type. We conclude in Section 5.

\clearpage
\section{Sample Selection and Preliminary Analysis}

\subsection{Sample Selection}

\begin{figure}
\epsscale{1.0}
\plotone{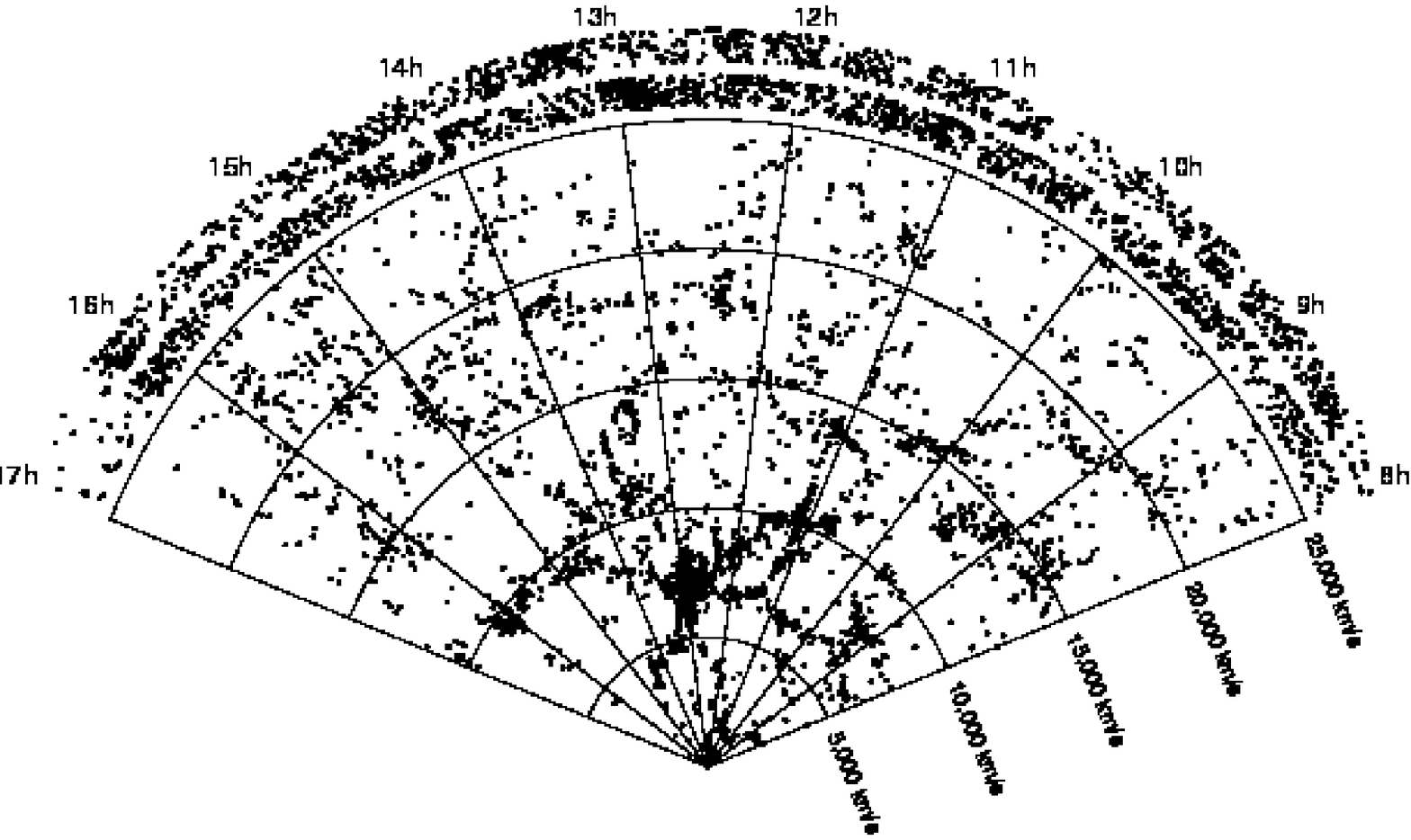}
\caption{The 15R sample in redshift space and in the
plane of the sky.  The cone diagram for $cz \leq 25,000$ km s$^{-1}$
superposes the two declination strips. 258 galaxies lie at greater
redshifts.
\label{fan}}
\end{figure}

We draw our galaxy sample from the northern portion of the 15R
Redshift Survey.  
\citep[The data for the entire survey will appear in][]{Geller}.
The 15R North Survey includes all galaxies with
Kron-Cousins \citep[see, e.g.,][]{Cousins,Bessell90,Bessell95}
$R\leq15.4$ and within two 2.5$^\circ$ strips
straddling the Century Survey \citep[CS hereafter]{Century}. 
The two
strips cover $8^h \le \alpha \le 17^h$ and $26.5^\circ \le \delta \le
29.0^\circ$ and $30.0^\circ \le \delta \le 32.5^\circ$ (B1950), a
total of $\sim 0.16$ steradians. Figure \ref{fan} shows the
distribution of the 15R North galaxies in the plane of the sky and in
redshift space.  Henceforth, we use 15R as a shorthand for 15R North,
and we use $R$ to denote Kron-Cousins magnitude.

\subsection{Photometric Calibration}

We obtained the initial 15R galaxy catalog from the POSS I E plates,
scanned at STScI with 25$\mu$ pixels; the catalog covers the entire
plate including both the CS and 15R strips.  Using techniques
described by \cite{McLean}, we reduced the data to instrumental
magnitudes and obtained star/galaxy classifications.  These initial
galaxy catalogs often omit the brightest objects.  We recovered
missing galaxies from the \cite{Zwicky} catalog.

The CS passes across the central portion of each 15R plate.  We
therefore calibrated each 15R plate by comparing the instrumental
magnitudes with the CS magnitudes \citep{Century}.  Geller et
al.~calibrated the photographic CS magnitudes against the CCD drift
scans of \cite{Kent, Ramella},
along with $\sim200$ pointed CCD measurements.  On each plate there
are $\sim 40$ galaxies with $R \leq 15.4$ in the CS region.  We use CS
magnitudes of these galaxies to calibrate the instrumental isophotal
15R magnitudes.  We fit each plate separately for both zero point and
slope, fitting the medians of the data in 0.5 mag bins.

For one plate, E1393, we used a CCD survey of the Coma cluster
\citep{Kashikawa95, Kashikawa98} to test our
photometric calibration. We have 120 galaxies in common with Kashikawa
et al.\ with $R \leq 15.4$.  The 1 $\sigma$ difference between the 15R
and Kashikawa et al.\ measurements is 0.18 mag for galaxies fainter
than $R = 14.0$. For galaxies with $R \leq 14$, the 15R measurements are
systematically too faint by 0.4 mag per mag; these comprise 7\% of the
survey.

\subsection{Spectra and Redshifts}

We obtained all of the spectra with the FAST spectrograph on the
Tillinghast 1.5m telescope on Mt.~Hopkins. We extracted each spectrum
from a single exposure of 180 to 600 s.  In all cases, the
spectrograph slit is 3 arcseconds wide and 3 arcminutes long, aligned
E-W.  We extract the raw 2D spectra (as described in $\S$ 2.6) and
produce 1D wavelength-calibrated spectra, with 3600---7400 {\rm \AA}
spectral coverage, 6 {\rm \AA} FWHM resolution, and typical S/N $\sim$
10.  This spectral range includes H$\alpha$ for all but three 15R
galaxies at $z$ $\sim$ 0.15.  Plate 1 is a color plate of all the 15R
spectra discussed in this paper, sorted by H$\alpha$ equivalent
width.

We make a correction for atmospheric water and O$_2$ absorption near
7000\AA\ in the $\sim$25\% of the cases where the absorption might
affect the measurement of H$\alpha$ 6563\AA\ or [N\II] 6548,6583\AA\
line strengths.  We derived the correction from several exposures of
the standard star CygOB2\#9, scaled by the effective airmass of
each exposure.

We use observations of Massey flux-standard stars \citep{Massey}
obtained during each observing run to flux calibrate the galaxy
spectra.  We derive a separate calibration for each run, typically 1
to 2 weeks long.  By comparing the flux calibrated spectra of galaxies
observed more than once, we estimate that the RMS error in the flux
calibration is 6\% over most of the spectral range, increasing to
$\sim$10\% below 4500 {\rm \AA}.

We measured equivalent widths of prominent emission features before
flux calibrating the spectra.  Table \ref{lines} lists the index
definitions of the relevant lines.  We use the convention that
positive equivalent widths denote emission.

\begin{figure}
\epsscale{1.0}
\plotone{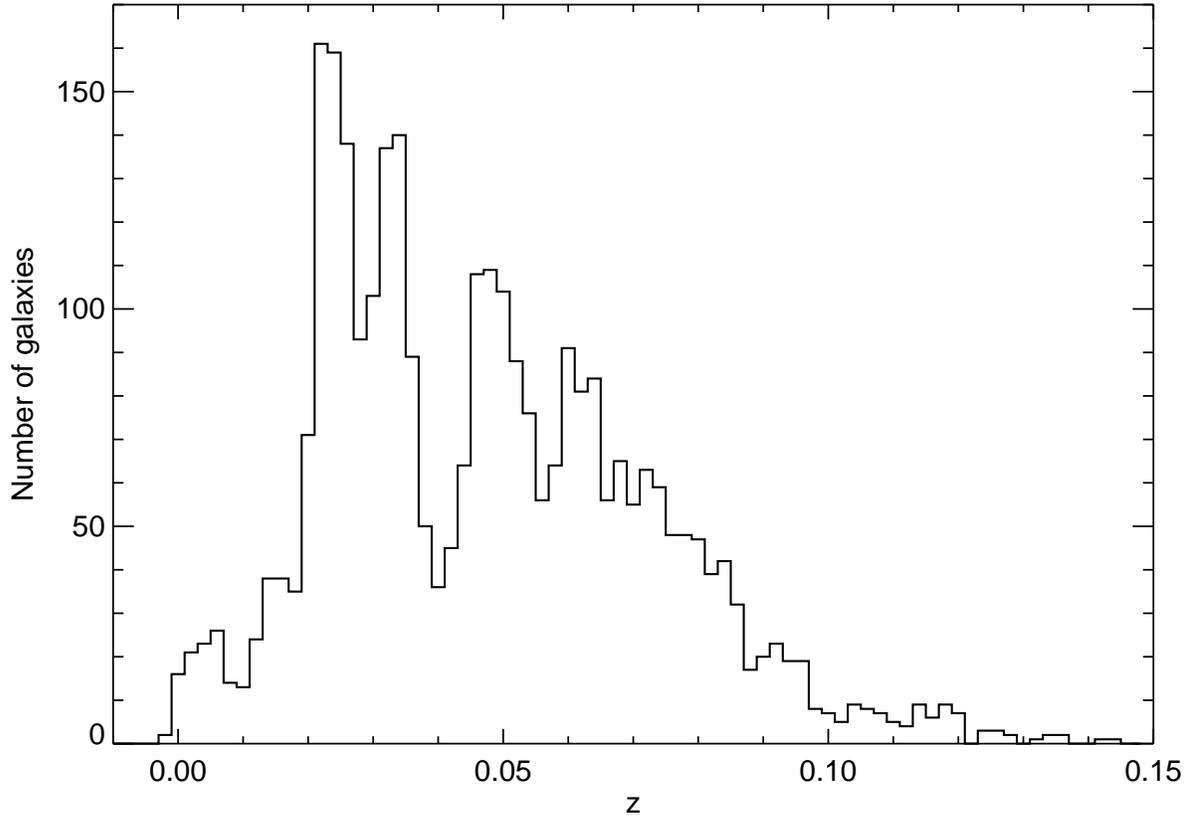}
\caption{Redshift histogram for the 15R sample.
\label{redshift}}
\end{figure}

We used the cross correlation techniques of \cite{Kurtz} to obtain
redshifts.  The average internal error in $cz$ reported by the XCSAO
routine is 35 km s$^{-1}$.  We extract heliocentric
redshifts, uncorrected for Virgo infall. Figure \ref{redshift} shows
the redshift of the sample; the median redshift is 0.05.

\placefigure{redshift}

\subsection{Absolute Magnitudes}

Given the apparent Cousins-R magnitude $R$, we apply a 
$k$-correction, $k(z)$, and Galactic reddening correction, $A_R$, 
to compute the absolute Cousins-R magnitude $M_R$:
\begin{equation}
M_R = R - 5 \log(D_{Mpc}) - 25 - A_R - k(z), \label{Mag}
\end{equation}
where the luminosity distance is
$$D_{Mpc} = \frac{c}{H_o q_o^2}\left[q_o z + 
            (1-q_o)\left(1-\sqrt{1+2q_o z}\right)\right].$$
We assume $H_o = 100$ km sec$^{-1}$ Mpc$^{-1}$ and 
$q_o = 0.5$.

\begin{figure}
\epsscale{1.0}
\plotone{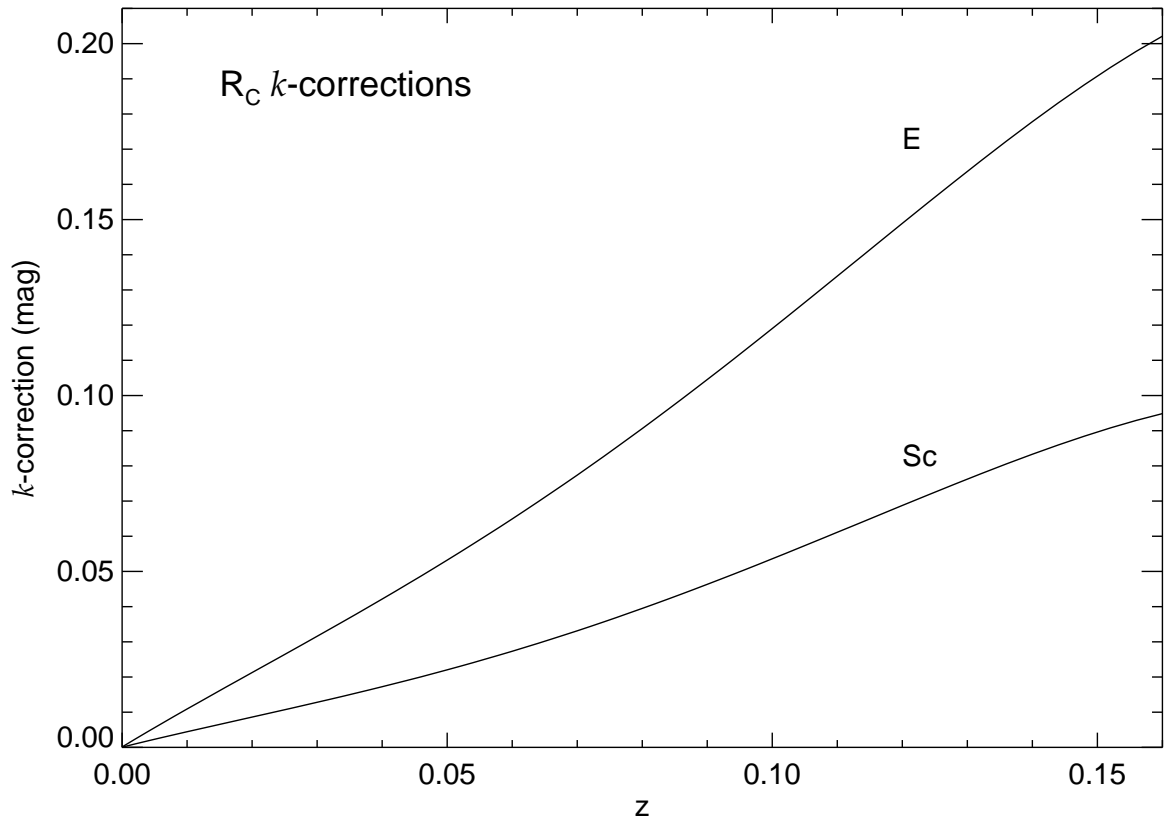}
\caption{Cousins-R $k$-corrections for
absorption line galaxies are an average of the ``E'' K-corrections
from \cite{Poggianti} and \cite{Frei}.  For emission-line galaxies we
average the ``Sbc'' and ``Scd'' K-corrections from \cite{Frei}
and the ``Sc'' K-correction from \cite{Poggianti}. 
\label{corr}}
\end{figure}

We use $k$-corrections from \cite{Poggianti} and \cite{Frei}.  We do
not correct for evolutionary effects because they are strongly
model-dependent, and are generally negligible at small redshifts.  The
largest $k$-correction is $0.18$ mag, less than the uncertainty in an
individual apparent magnitude.  We apply an elliptical $k$-correction
for galaxies with absorption-line spectra and an Sc $k$-correction
function for galaxies with emission-line spectra ( $\S$3 gives our
definitions of absorption-line and emission line spectra).  We average
the elliptical galaxy $k$-corrections of \cite{Poggianti} and
\cite{Frei}.  For the emission-line galaxies we average the
\cite{Frei} Sbc and Scd and the \cite{Poggianti} Sc $k$-corrections.
Figure \ref{corr} shows the adopted $k$ corrections.

\placefigure{corr}

For Galactic extinction, we follow \cite{Winkler97}:

$${{A_R} \over {E(B-V)}} = {{A_V} \over {E(B-V)}} - 0.6$$

Here $A_{X}$ is the extinction in magnitudes in the filter passband
$X$, and the reddening, $E(B-V)$ is $A_B - A_V$.  Following
\cite{Savage79} we adopt:

$${{A_V} \over {E(B-V)}} \sim 3.~~~ {\rm Then:~~} 
{{A_R} \over {E(B-V)}} \sim 2.4. $$

We derive the reddening from the relation of \cite{Bohlin}:

$$E(B-V) = \frac{N_{HI}}{4.8\times10^{21}\ \mbox{cm}^{-2}}$$ where
$N_{HI}$ is the column density of neutral hydrogen in atoms cm$^{-2}$
\citep{Burstein}.  For the 15R sample, the median value of $A_R$ is
0.09 magnitudes, with extremes of 0.04 and 0.24 mag.

\begin{figure}
\epsscale{1.0}
\plotone{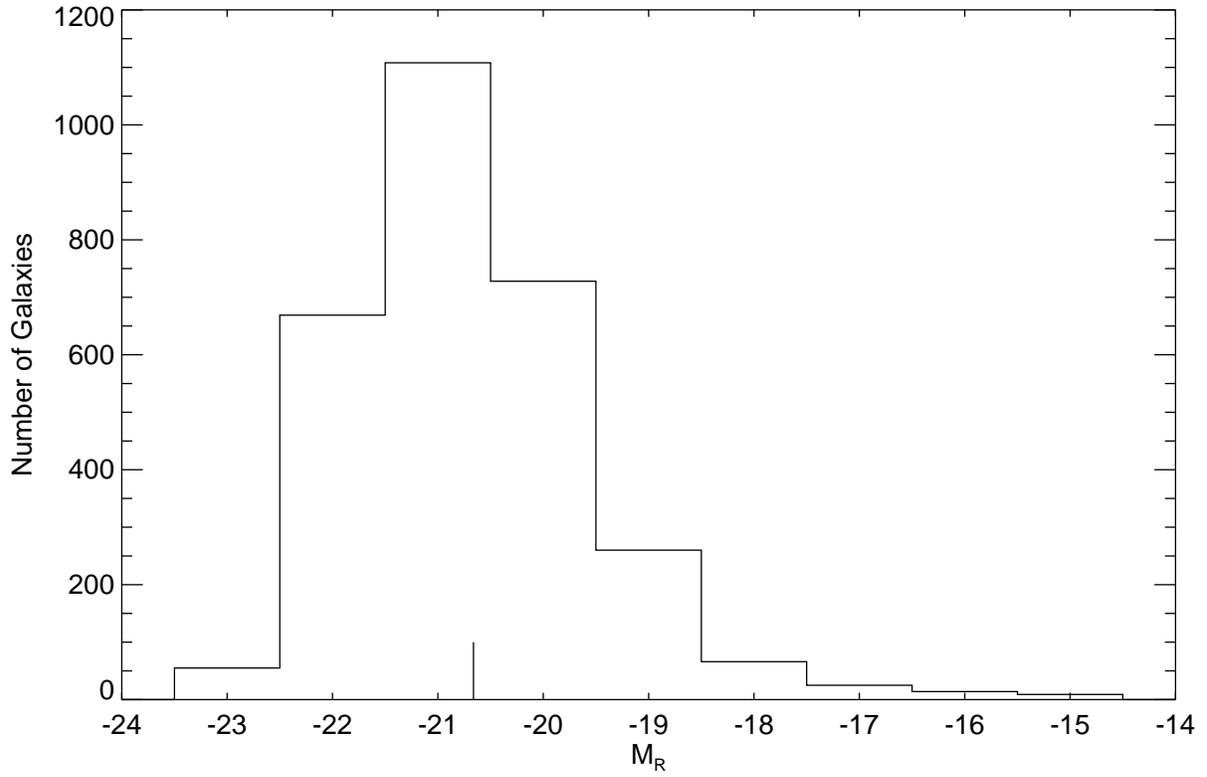}
\caption{Histogram of absolute R magnitudes.
The mean M$_R = -20.7$.
\label{mags}}
\end{figure}

Figure \ref{mags} shows the distribution of the corrected absolute
magnitudes, $M_R$.  The mean 15R $M_R$ of -20.7 is close to the
Century Survey $M_\ast(R) = -20.73$ \citep{Century}.

\placefigure{mags}

\subsection{Slit Covering Fraction}
\label{ap_sec}

The emission line equivalent widths we measure may depend on the
fraction of the integrated galaxy flux that falls on our slit.  
\citet{Kochanek} demonstrate that small slit (aperture) covering fractions 
may introduce a bias in the estimate of emission line equivalent widths. 
They argue that in fiber surveys the small covering fraction generally
leads to an underestimate of the equivalent width and a corresponding
overestimate of the fraction of galaxies with absorption-line spectra
only. The bias depends on the morphological type of the galaxy; the
larger the bulge to disk ratio, the more serious the potential bias
under the assumption that the emission lines originate primarily in
the disk.

We collect nearly all of the nuclear light, but the fraction of light
beyond that depends on the redshift, intrinsic size, position angle,
and surface brightness of the galaxy.  The slit has a physical size of
2.2 kpc by 132 kpc (maximum) at the survey median redshift $z=0.05$.

\begin{figure}
\epsscale{1.0}
\plotone{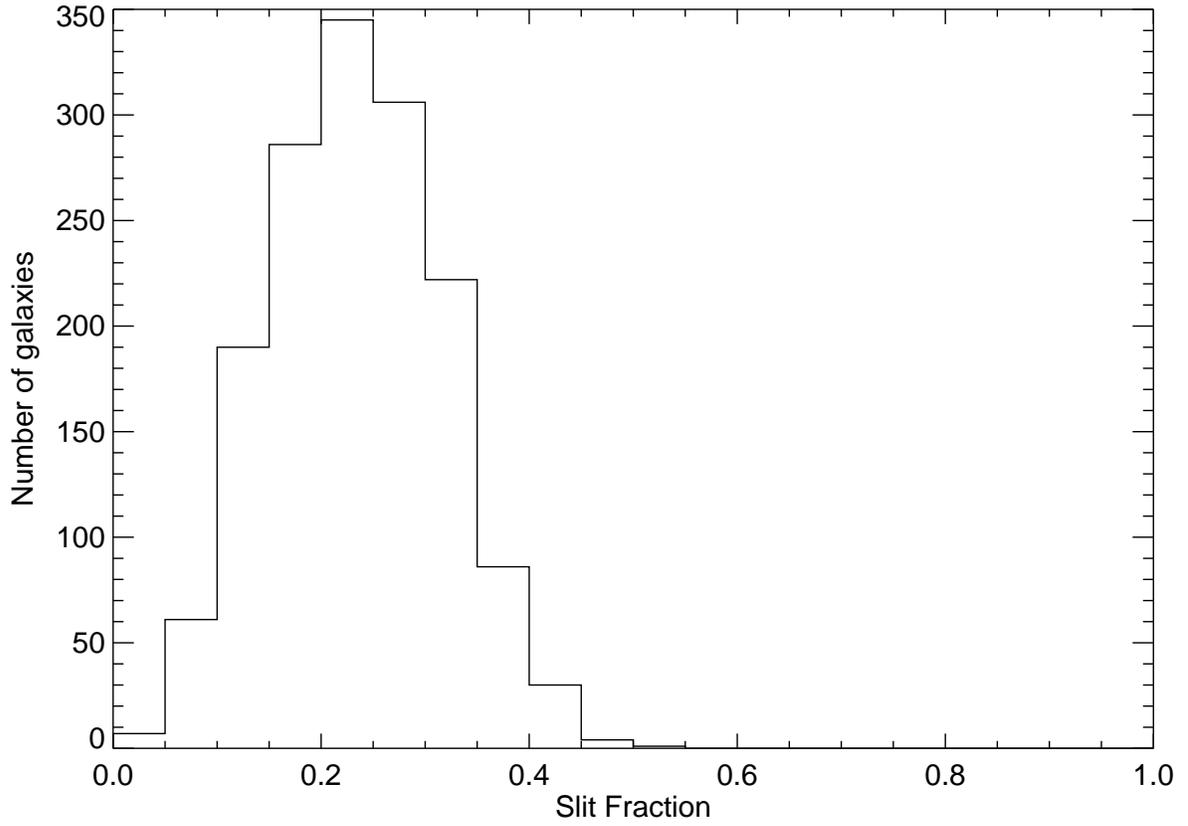}
\caption{Fraction of the total galaxy flux gathered by the spectrograph
slit for emission-line galaxies, as estimated from the digitized
POSS images.  The median slit fraction is 0.24.
\label{slit}}
\end{figure}

We use the digitized Palomar (POSS) images of the 15R galaxies to
estimate the slit covering fraction for the emission-line galaxies.
We measure the ratio of the photographic flux within the slit aperture
(centered on the nucleus) to the total photographic flux.  We do not
correct for the photographic nonlinearity inherent in the POSS
images. Figure \ref{slit} shows that most of the 15R-north spectra
contain 15---25\% of the total flux of the galaxy; for the nearest
galaxies this fraction can be as small as 2\%.  The uncertainty in
these corrections is $\sim$10\% of the correction.

\placefigure{slit}

\begin{figure}
\epsscale{.8}
\plotone{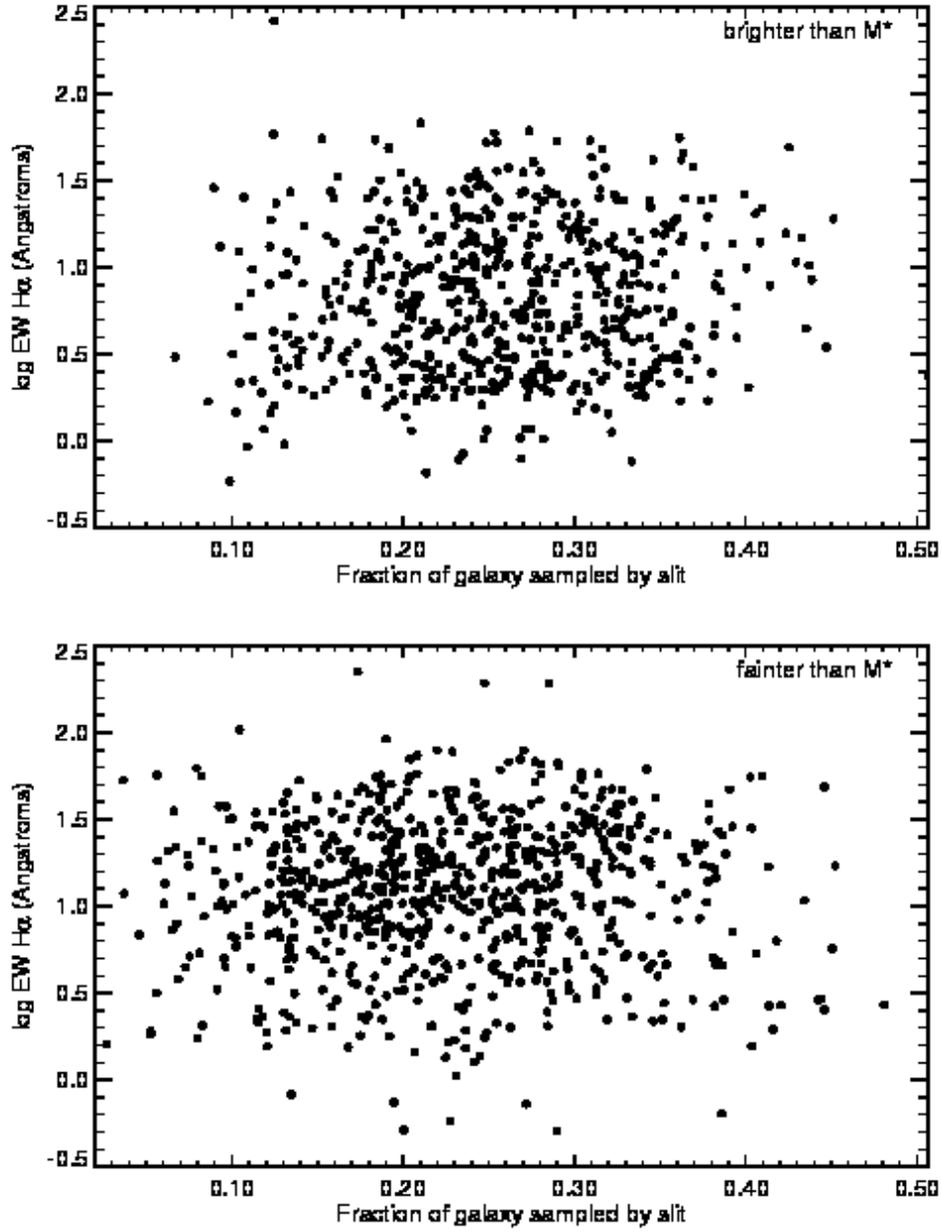}
\caption{H$\alpha$ equivalent width as a function of slit fraction.
\label{Ha-slit}}
\end{figure}

For emission-line galaxies (see $\S$ \ref{classification} for the
quantitative discriminant we use to distinguish emission-line and
absorption-line galaxies), Figure \ref{Ha-slit} shows the
EW(H$\alpha$) as a function of slit covering fraction. There is no
trend of EW(H$\alpha$) with slit covering fraction for galaxies either
brighter than or fainter than L$^*$.

\begin{figure}
\epsscale{.8}
\plotone{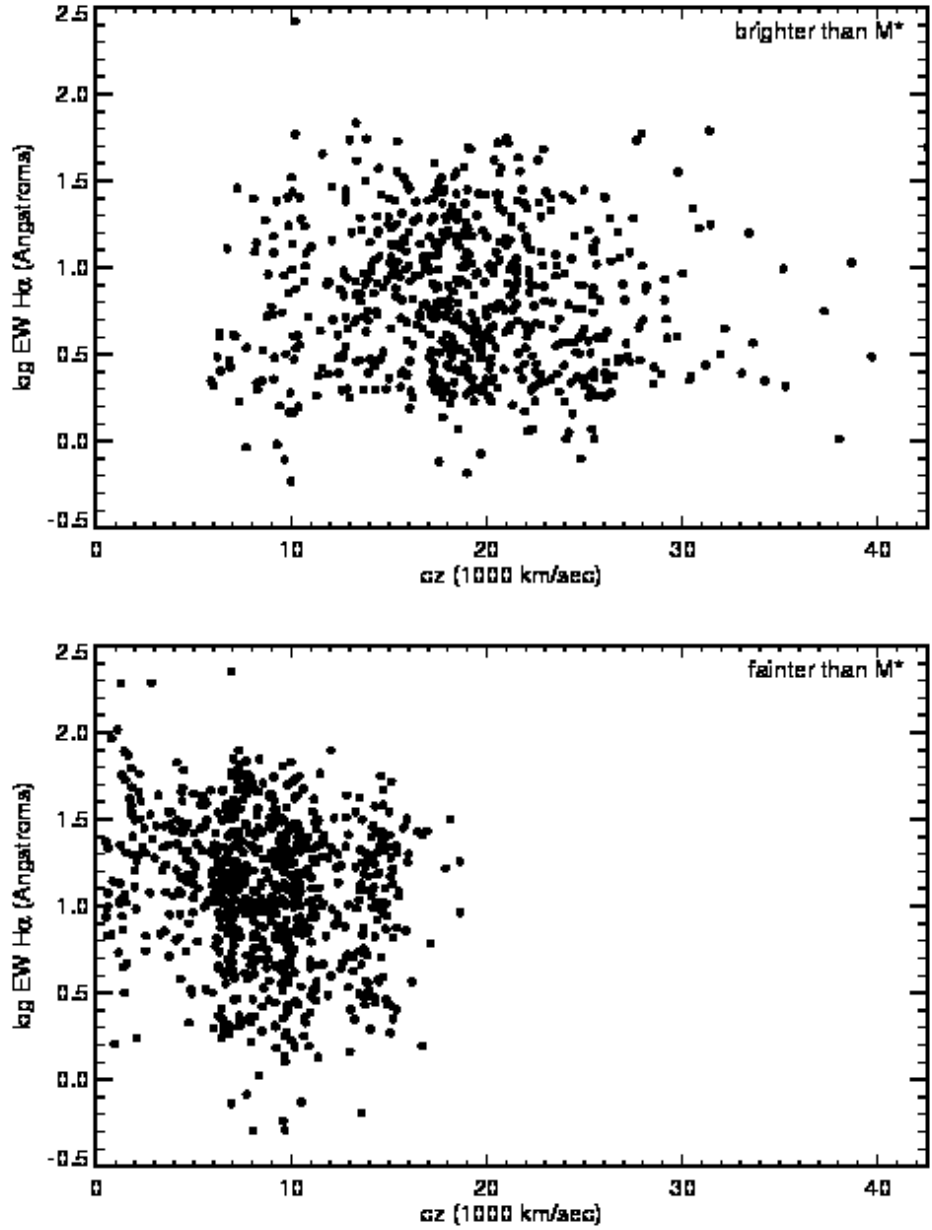}
\caption{H$\alpha$ equivalent width as a function of redshift.
\label{Ha-cz}}
\end{figure}

Figure \ref{Ha-cz} shows the EW(H$\alpha$) as a function of $cz$ for
intrinsically bright and faint galaxies. For galaxies with $L\geq
L^*$, there is no trend of EW(H$\alpha$) with $cz$; for galaxies with
$L < L^*$, there is a decline in EW(H$\alpha$) with $cz$. The latter
dependence is opposite to the bias discussed by \citet{Kochanek}; it
occurs because actively star-forming dwarf irregulars are within the
survey limiting magnitude only at very low redshift. We conclude that
the slit covering fraction does not introduce any substantial bias in
our evaluation of emission line equivalent widths.

\placefigure{Ha-slit}
\placefigure{Ha-cz}


\clearpage
\section{Spectral Classification and Star Formation}

To study star formation and to examine the dependence of spectral type
on the local environment, we classify the spectral type of the 15R
galaxies.  We use ratios of strong emission lines (H$\alpha$, [N\II]
6583, [O\III] 5007, and H$\beta$) to separate star-forming galaxies
from galaxies containing a strong AGN.  Because the 15R spectra
include both H$\alpha$ and [O\II] 3727, we can also assess [O\II] as a
measure of the star formation rate.

\subsection{Spectral Classification}
\label{classification}

We use emission line indices (see Table \ref{lines}) to measure the
emission line equivalent widths.  We first separate the spectra into
two broad categories, emission and absorption.  Plate 1 shows that
roughly half of the spectra have detectable emission lines.  We
classify a galaxy as an emission-line galaxy if we detect one or more
emission lines (Table \ref{lines}) at $\geq 2 \sigma$.  If no single
emission line meets this criterion, we still classify the spectrum as
an emission-line spectrum if the emission detections add in quadrature
to at least $2 \sigma$.  1538 of the 3149 15R galaxies have
emission-line spectra.  We classify the remaining spectra as
absorption spectra; stellar absorption lines are, in fact, present in
all of the spectra (Plate 1).

\begin{figure}
\plotone{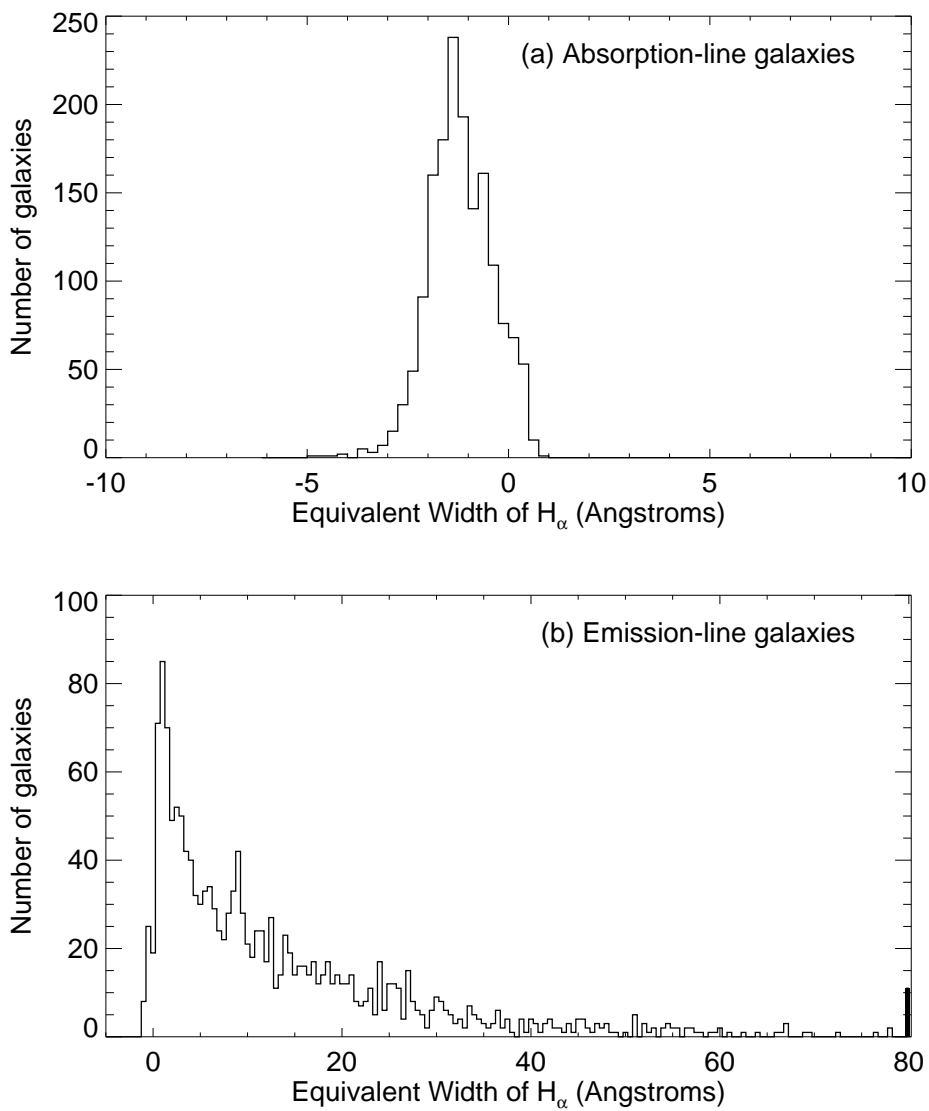}
\epsscale{.9}
\caption{Histograms of the H$\alpha$ equivalent widths
for all the galaxies in the 15R sample: (a) absorption line galaxies
and (b) emission line galaxies. Positive values indicate emission.  We
classify galaxies with H$\alpha$, H$\beta$, [N\II] 6583, or [O\III]
5007 emission lines adding up to at least 2$\sigma$ as emission-line
objects; we detect no H$\alpha$ emission in a few of these.  The last
bin in the emission line panel contains galaxies with EW(H$\alpha$)
$\ge$ 80{\rm \AA}.
\label{Ha-histo}}
\end{figure}

Figure \ref{Ha-histo} shows the distribution of the EW(H$\alpha$)
for the entire sample.  The median uncertainty for an individual
equivalent width is $\sim$0.5{\rm \AA}.  The 49\% of 15R galaxies with
emission-line spectra is less than the 61\% with EW(H$\alpha$) $>$
2{\rm \AA} detected in the $b_j$ selected Stromlo-APM Survey \citep{Tresse}.
The greater detection rate of emission line galaxies in the Stromlo-APM
survey probably results from their $b_j$ rather than $R$
selection.  Their wider slit, which collects a larger fraction of the
disk light, may also contribute somewhat to this difference.

\placefigure{Ha-histo}

The modest S/N of the 15R spectra complicates the classification of the
absorption-line galaxies and we defer that to a later work.  We
classify the emission-line galaxies according to the line-ratio scheme
described by \citet{BPT} and \citet{VO} (VO hereafter).  
For this classification, we use line fluxes measured from
the flux-calibrated spectra, and we correct for 1.54 \AA\ in
equivalent width of stellar Balmer absorption at H$\alpha$, based on
high S/N measurements from a smaller but statistically similar sample
(R. Jansen 1999, private communication).

\begin{figure}
\epsscale{0.5}
\plotone{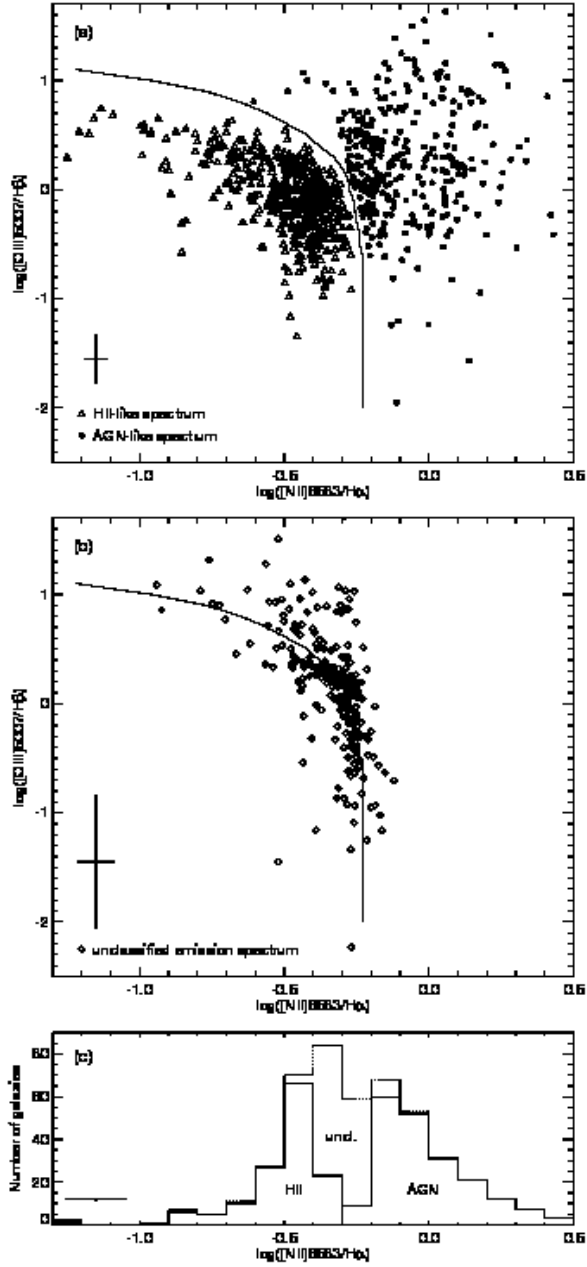}
\caption{Classification of the emission-line galaxies according to the
line-ratio method of \cite{VO}.  (a) \& (b) show those galaxies with
all four emission lines, (c) those missing [O\II] and/or H$\beta$.
(a) The unambiguous H\II-like or AGN-like galaxies have at
least 90\% of the error ellipse area on the appropriate side
of the dividing line.
(b) Objects that overlap the dividing line 10-90\% are left unclassified.
(c) Objects lacking either [O\III] or H$\beta$ in
emission are classified according to their [N\II]/H$\alpha$ ratio
alone, again with 90\% error-bar overlap required for unambiguous
classification. 
Indicative error bars (lower left) show the median errors in the three cases.
\label{vo}}
\end{figure}

Figure \ref{vo} (a \& b) plots the line flux ratios of
[O\III] 5007\AA\ / H$\beta$ and [N\II] 6583\AA\ / H$\alpha$.  The
plots also show the empirical relation derived by VO to separate AGNs
from normal star-forming galaxies.  The 15R galaxies do not divide
cleanly into two separate populations.  Higher S/N spectra would
obviously determine how much our measurement errors contribute to this
blur between AGN and star-forming galaxies. 
The hybrid spectra (Figure \ref{vo}(b)) may result from both 
observational and astrophysical sources. 
On the observational side, our longslit spectra blend nuclear and disk light. 
Extraction of nuclear spectra, a task beyond the scope of this
paper, could clarify this issue. From the broad astrophysical
perspective, many groups have examined the AGN-starburst connection
\citep[e.g.,]{Kewley, Lei, Coziol}.  \cite{Weedman} and \cite{Norman}
suggest that nuclear starbursts may lead to the formation of a massive
central black hole, and, conversely, the presence of an AGN in a
galaxy may trigger a burst of star formation \citep{Rees, Daly}. 
\cite{Donzelli} examine a set of merging galaxy pairs and find
that some galaxies host a low-luminosity AGN surrounded by strong
star-forming regions. Evaluation of the frequency of a true hybrid
AGN-starburst population in the 15R survey requires a more
sophisticated analysis of the data than that discussed here.

In some cases, we do not detect the four emission lines required by
the VO classification scheme in the 15R spectra.  When at least
H$\alpha$ and [N\II] 6583\AA\ are present, we use the ratio of these
two lines for the classification.  If $\log$([N\II]6583/H$\alpha$) $<$
$-0.25$ we classify the galaxy as star forming (H\II-like); otherwise
we classify it as an AGN\@.  Figure \ref{vo}(c) shows histograms of the
galaxies classified using only the H$\alpha$ and [N\II] 6583\AA\
emission lines.

\placefigure{vo}

\begin{figure}
\epsscale{0.7}
\plotone{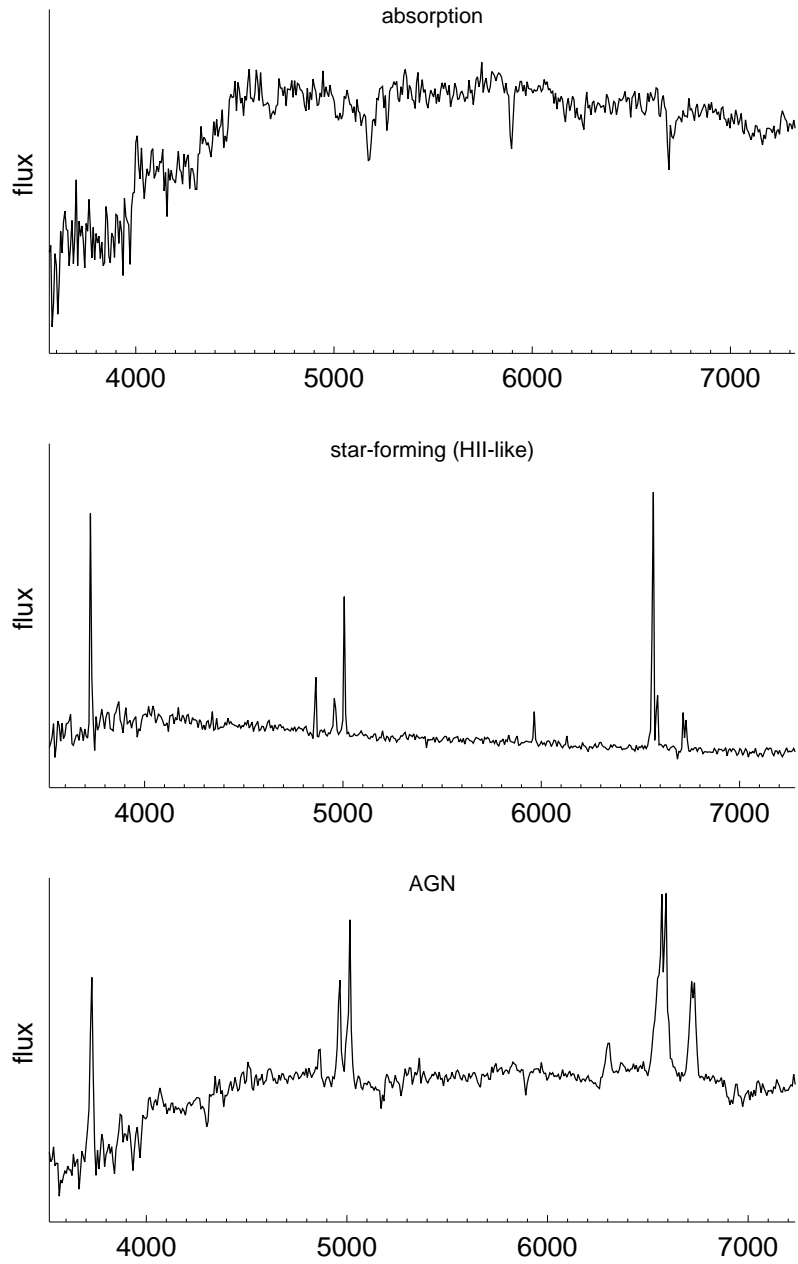}
\caption{Sample spectra of galaxies classified as absorption line, star-forming
(H\II-like), and AGN.
\label{example1}}
\end{figure}

\begin{figure}
\epsscale{0.7}
\plotone{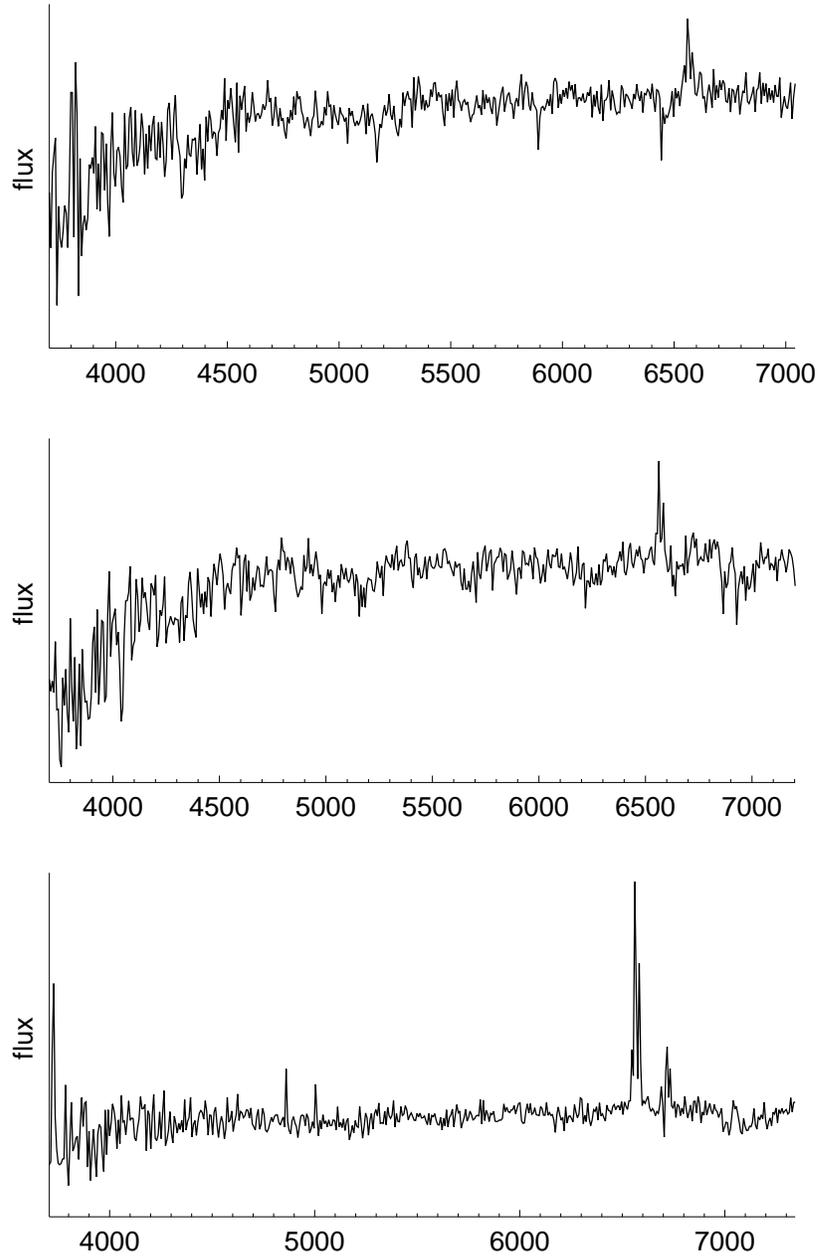}
\caption{Three spectra of galaxies with unclassified emission.
\label{example2}}
\end{figure}

The error bars for many of the emission line ratios (both those with
all 4 lines and those using only [N\II]6583/H$\alpha$) often overlap
the VO dividing line.  We classify each emission-line galaxy according
to the area covered by its error ellipse.  We assign the ``H\II''
designation to each galaxy with less than 10\% of its error ellipse on
the AGN side, and ``AGN'' to galaxies with less than 10\% on the H\II\
side.  Galaxies overlapping the VO line by 10 to 90\% remain
unclassified; these constitute $\sim$25\% of the emission-line
galaxies.  Synthetic rest-frame $(B-V)_0$ colors confirm the
classification of H\II-like and AGN-like spectra.  As expected, the
H\II-like spectra tend to be bluer than the AGN-like spectra.  The
unclassified spectra tend to be closer in color to the AGN-type
spectra. Figures \ref{example1} and \ref{example2} show examples of
absorption-line, AGN, star forming, and unclassified emission-line
spectra.

\placefigure{example1}
\placefigure{example2}

Table \ref{types} lists the numbers of each spectral type in the 15R
sample.  We classify 17\% of the total sample as AGN; given the
unclassified emission-line spectra this fraction is a lower limit.
Our AGN fraction is much larger than the $<$ 1\% of galaxies
classified as AGNs in the Las Campanas Redshift Survey (LCRS)\@.
Because the LCRS spectra do not usually include H$\alpha$, their
classifications are based on the equivalent widths of [O\II] 3727\AA,
[O\III] 5007\AA, H$\beta$ 4861\AA, and [Ne\V] 3425\AA\ \citep{LCRS}.
The LCRS procedure is less reliable than the VO technique we use.  We
conclude that a significant number of low-luminosity AGNs are probably
classified as star-forming galaxies in the LCRS.

\begin{figure}
\epsscale{1.0}
\plotone{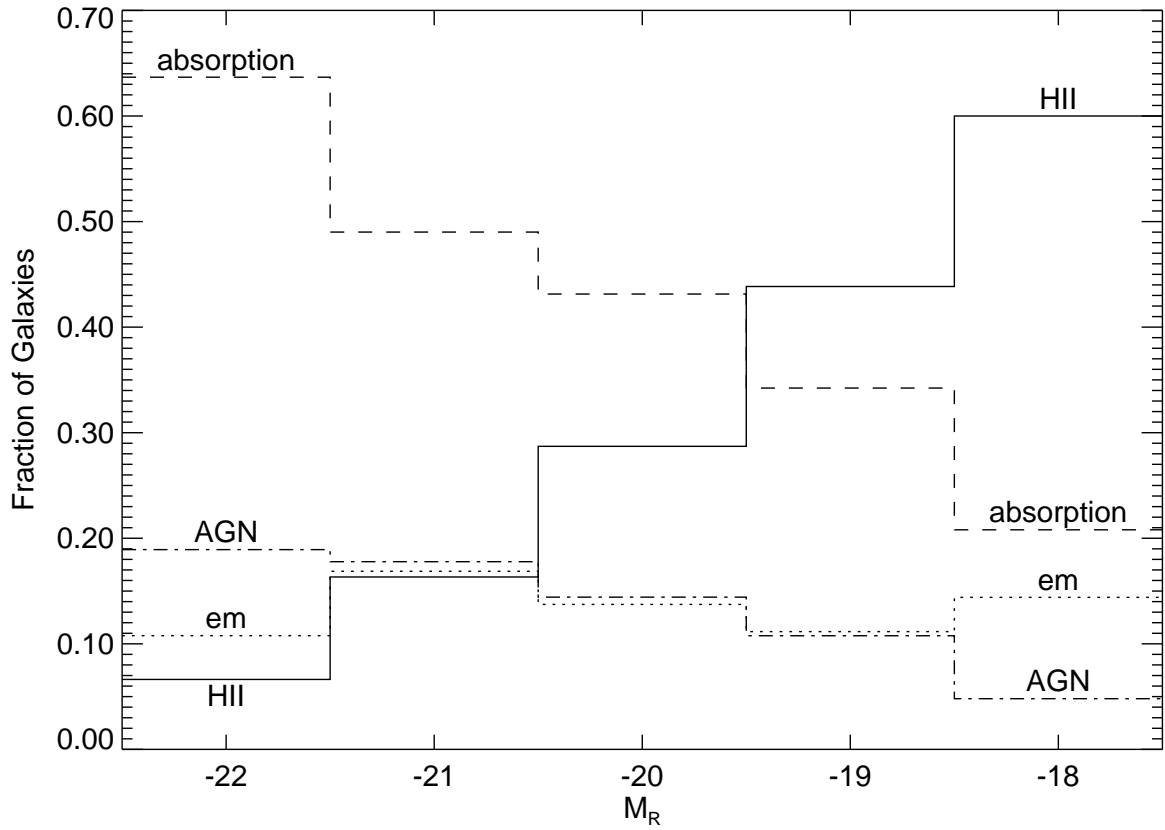}
\caption{Spectral types as a function of absolute R magnitude.  
The four types are absorption-line, AGN, star-forming (H\II), and 
unclassified emission (``em'').  
As expected, the most luminous galaxies are largely absorption-line
galaxies; the least luminous are star-forming (H\II).
\label{histo1}}
\end{figure}

Figure \ref{histo1} shows the relative fractions of the galaxy spectral 
types as a function of absolute magnitude.  
The absorption-line fraction declines steeply with decreasing luminosity,
and the emission line fraction increases.  
AGNs are more abundant among more luminous galaxies.
The absorption-line galaxies are significantly more luminous than the 
star-forming galaxies.  
The median $M_R$ of the absorption-line, AGN, unclassified emission-line, 
and star forming galaxies are: -21.1, -21.1, -20.7, and -20.1, respectively.

\placefigure{histo1}

\subsection{The Spatial Distribution of Different Spectral Types}

\begin{figure}
\epsscale{1.0}
\plotone{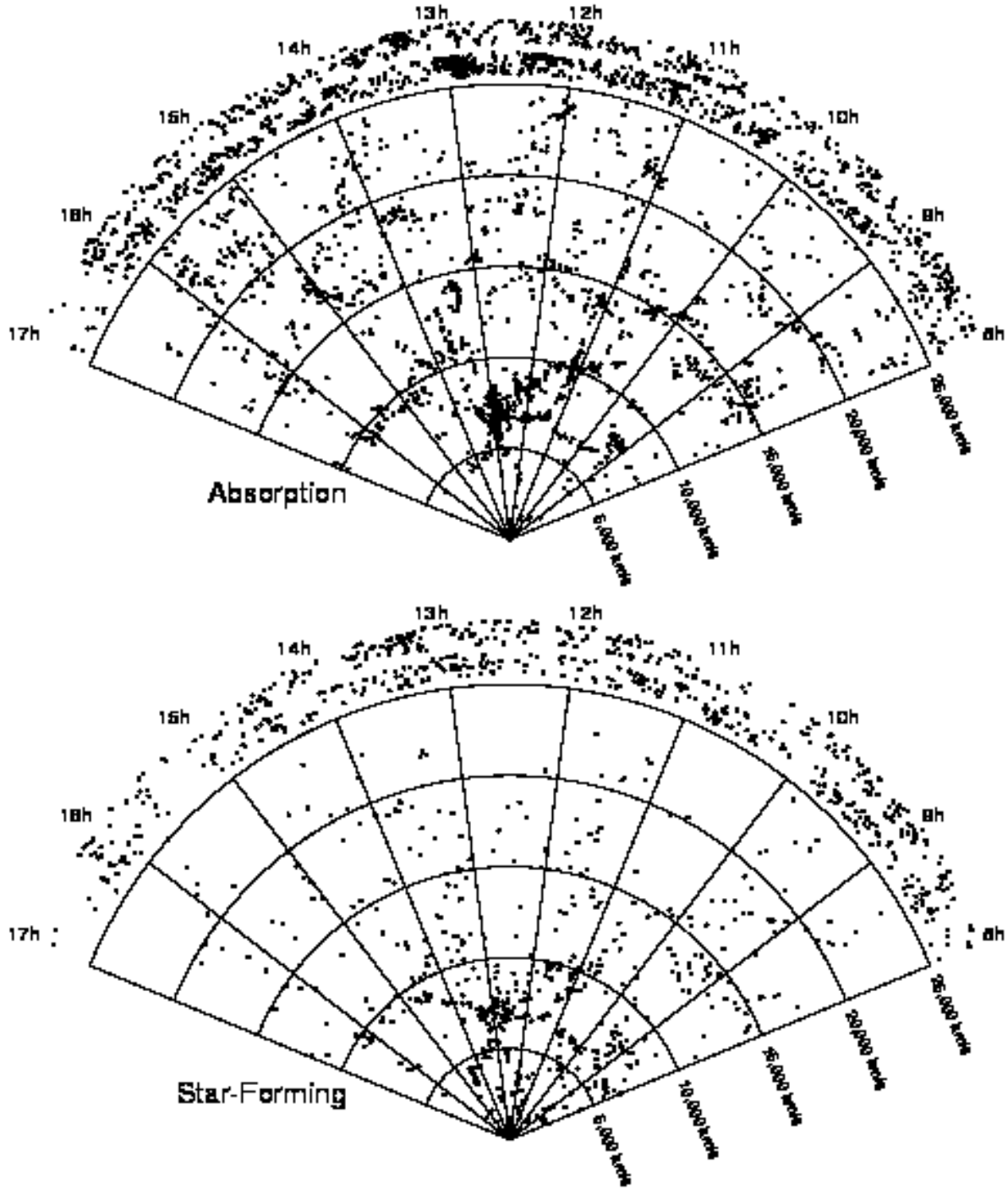}
\caption{The absorption line and star-forming 
subsamples in redshift space and in the plane of the sky (see Figure
\ref{fan} for the complete sample).  Absorption-line galaxies dominate
the Coma Cluster (the dense ``finger'' near the center of the
distribution). Star-forming galaxies appear in the infall region.
\label{fan_types_1}}
\end{figure}

\begin{figure}
\epsscale{1.0}
\plotone{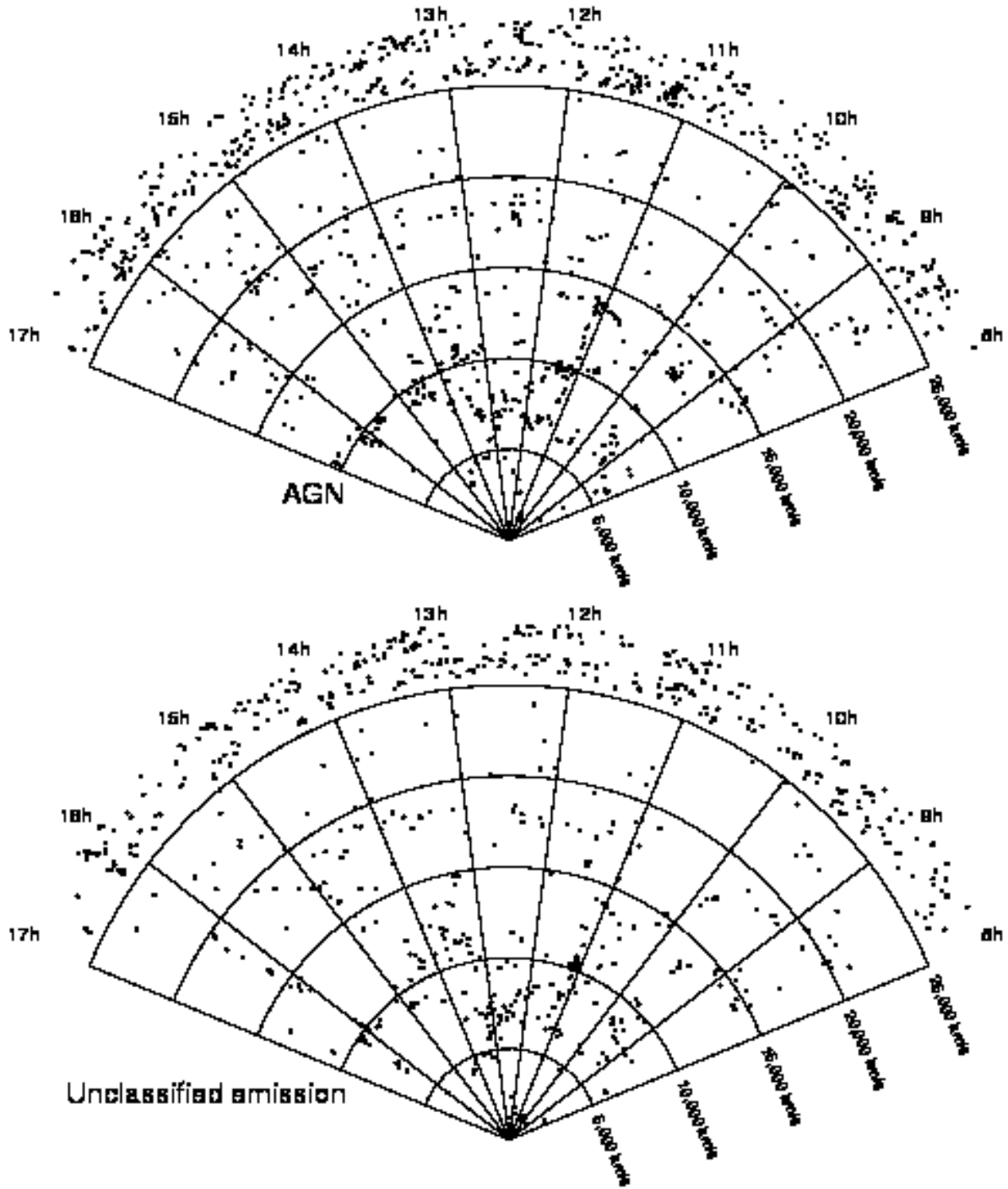}
\caption{The AGN and unclassified emission subsamples in redshift space and in
the plane of the sky (see Figure \ref{fan} for the complete sample).
The AGNs and unclassified emission-line galaxies trace the large scale
structure, but the contrast of the Coma Cluster is lower.
\label{fan_types_2}}
\end{figure}

Figures \ref{fan_types_1} and \ref{fan_types_2} show the distribution
of galaxies of the various spectral types in redshift space. As in
other surveys, the absorption-line galaxies are more strongly
clustered than the emission-line systems, and are particularly
abundant in the Coma cluster.  The AGNs and unclassified emission-line
galaxies trace the overall large-scale structure.  A number of H\II\
galaxies appear in the infall region around the Coma cluster. However,
they are generally absent from the central ``fingers'' which define
the virialized central regions of rich clusters. In Section
\ref{st_density} we return to this issue and compute the abundance of
each spectral type as a function of local density; overall the
relative fraction of H\II\ galaxies drops substantially in the densest
regions.

\placefigure{fan_types_1}
\placefigure{fan_types_2}

\subsection{H$\alpha$ + [N {\scshape~II\@}] and the Stellar Birthrate}

\cite{Kennicutt94} describe several techniques for estimating the
Scalo stellar birthrate parameter $b$, defined as the ratio of the
current star formation rate (SFR) to the time averaged SFR:
$$ b \equiv {{SFR} \over {< SFR >}} = {{SFR \cdot {\tau}_d} \over
{M_d}}(1-R_f),$$ where $M_d$ is the stellar mass, $\tau_d$ is the age
of the stellar population, and $R_f$ is the fraction of mass in each
stellar generation that is not tied up in stars \citep{Tinsley}.  The
current star formation rate can be determined directly from the
H$\alpha$ luminosity, and the stellar mass can be estimated from the
lumosity and mass-to-light ratio.  \citet{Kennicutt94} (hereafter KTC94)
adopt $R_f$=0.4, based on stellar evolution models, and take
$\tau_d \sim 10$ Gyr.  For their sample of 210 galaxies, they compare
this relatively direct estimate of the birthrate parameter $b$ to two
others based on the EW(H$\alpha$ + [N\II]) or $B-V$ colors predicted
by their stellar evolution and spectral synthesis models.  The $b$
estimated from the EW(H$\alpha$ + [N\II]) agrees well with the more
direct technique, but the estimate from the $B-V$ color has very large
scatter.  We therefore use the calibration (exponential + burst model)
from Figure 3 of KTC94 to convert our EW(H$\alpha$ + [N\II])
measurements to an estimate of $b$. KTC94 derive their estimates of
$b$ from integrated spectra. The absence of a dependence of
EW(H$\alpha$) on the slit covering fraction (Figure \ref{Ha-slit},
Section 2.5) indicates that our long-slit spectra should introduce
only small differences in the estimated $b$.

\begin{figure}
\epsscale{0.7}
\plotone{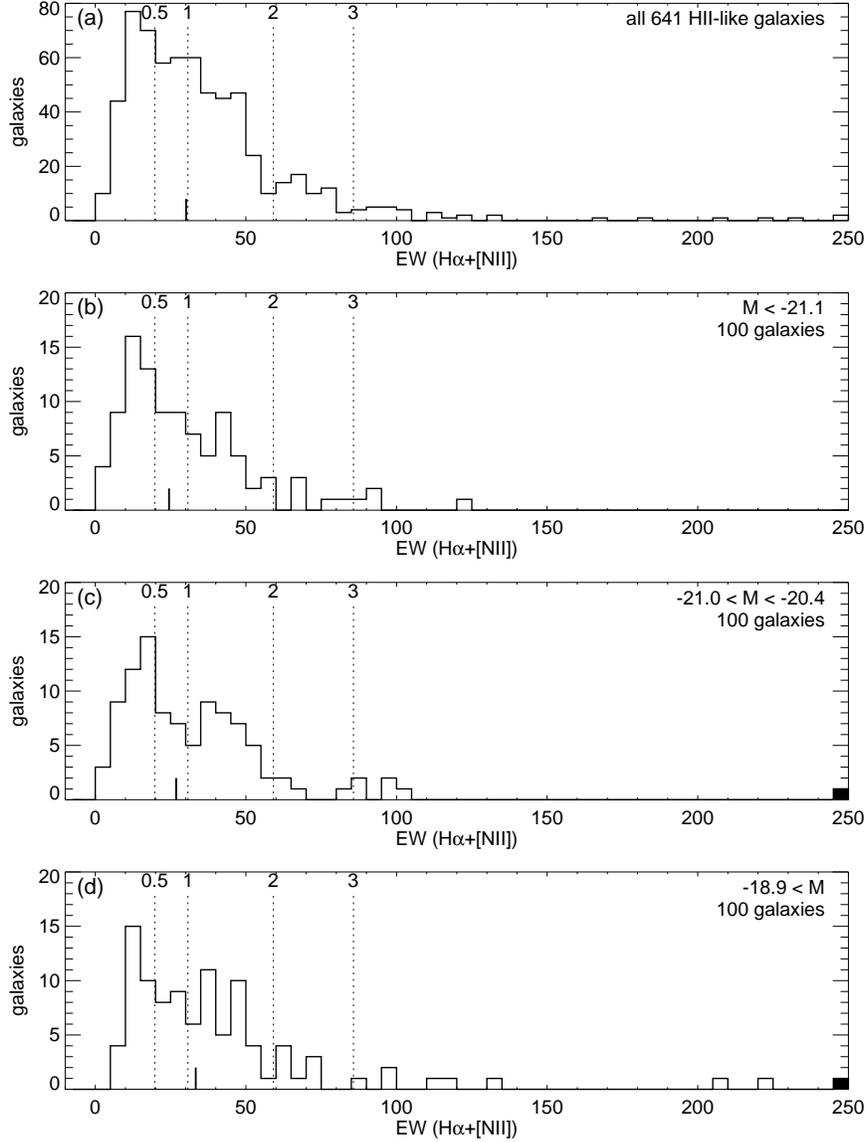} 
\caption{Histograms of the EW(H$\alpha$ + [N\II]) for
(a) the complete sample of star-forming (HII-like) galaxies, (b) the
100 most luminous star-forming galaxies, (c) 100 star-forming galaxies
closest in luminosity to M$_{\ast} = -20.7$, and (d) the 100 faintest
star-forming galaxies.  Dotted vertical lines indicate values of the
Scalo birthrate parameter, $b$, computed from the EW(H$\alpha$ +
[N\II]) following \cite{Kennicutt94}.  A short vertical bar marks the
median birthrate for each subsample. A solid box in the rightmost bin
in any panel contains galaxies with EWs $\ge$ 250{\rm \AA}.
\label{birthrate}}
\end{figure}

The stellar birthrate in KTC94 is greater on
average for later morphological types which have a lower average
luminosity.  In Figure \ref{birthrate} we plot the EW(H$\alpha$ + [N\II])
histograms for the complete sample of 641 star forming galaxies.
We also show the histograms for three 100 galaxy subsamples selected
by luminosity: the 100 most luminous galaxies, 100 galaxies with
luminosities near M$_*$, and the 100 least luminous galaxies,
respectively. In Figure \ref{birthrate}, vertical dotted lines mark the
approximate stellar birthrates from the KTC94 calibration.  
In agreement with KTC94, the median birthrate parameter is inversely
correlated with luminosity, increasing from 0.7 in the highest
luminosity subsample to 1.1 in the lowest luminosity subsample.

\placefigure{birthrate}

The largest birthrates in the KTC94 sample of 210 ``normal'' galaxies
are for 6 galaxies with $b \sim 3$.  Galaxies with $b \ge 3$ are
undergoing unusually vigorous star formation. KTC94 probably exclude
most of these galaxies from their sample by eliminating peculiar and
starburst galaxies.  In our complete sample of 641 star-forming
galaxies in 15R, 33 ($\sim$5\%) have $b > 3$. Table \ref{vigorous-sf}
lists these objects.  The samples are small, but the fraction of
galaxies with $b >3$ also appears to be inversely correlated with
luminosity.

\subsection{Using [O{\scshape~II\@}] to Estimate Star Formation}

\begin{figure}
\epsscale{1.0}
\plotone{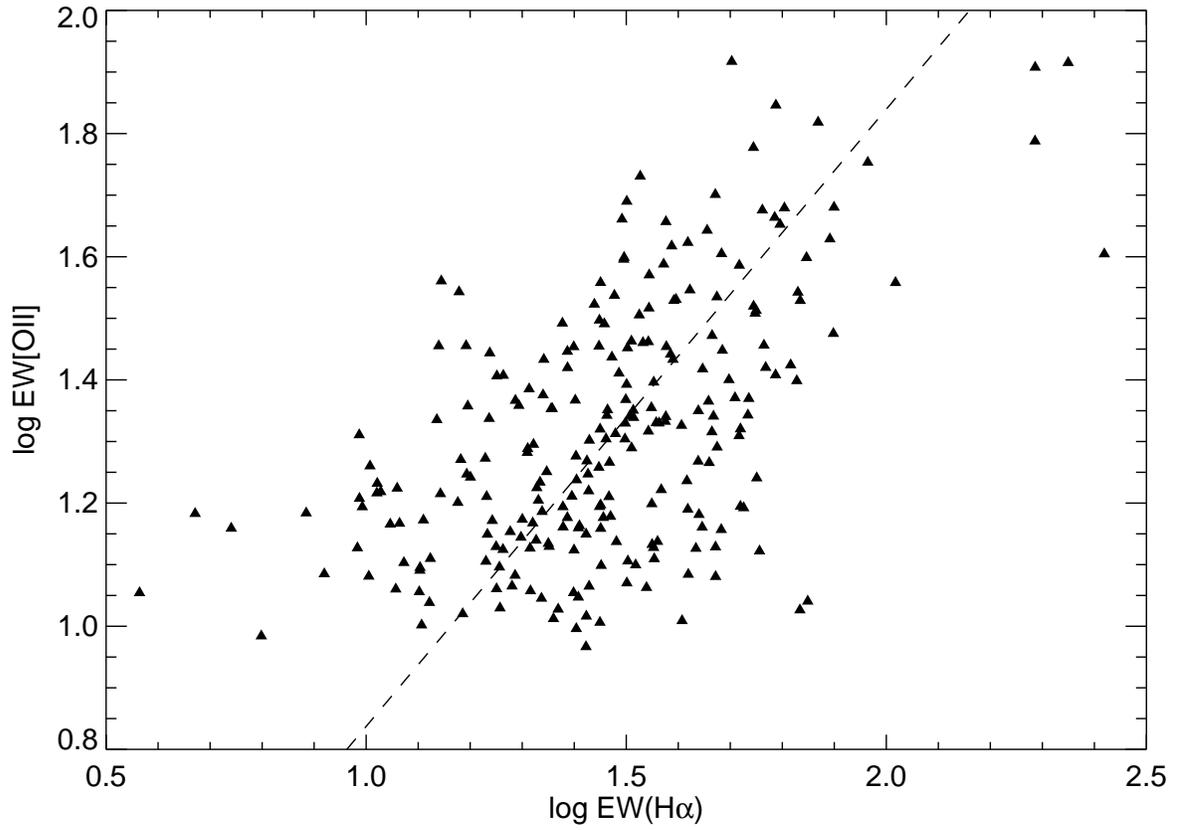}
\caption{Relation between [O\II] and H$\alpha$ EWs for
the 244 star-forming galaxies with [O\II] detected at $\ge$
2$\sigma$.  The dashed line shows the relation fit by \cite{Tresse}
to their data ([O\II] = 0.7H$\alpha$) which also fits our data well.
\label{OII_and_Ha}}
\end{figure}

Of the optical emission lines H$\alpha$ provides the most direct
measure of the star formation rate in a galaxy, but at intermediate
redshifts H$\alpha$ is outside of the optical bandpass.  The
strong [O\II] 3727\AA\ doublet is often a substitute measure of
star formation. \cite{Gallagher} and \cite{Kennicutt92}  show
that the line strengths of [O\II] 3727\AA\ and H$\alpha$ are
correlated.  \cite{Tresse}  recently confirmed the correlation
of EW([O\II]) and EW(H$\alpha$); EW([O\II]) $\sim$ 0.7
EW(H$\alpha$).  The \cite{Tresse} relation fits our data well
(Figure \ref{OII_and_Ha}). We plot log(EW([O\II])) vs.~log(EW(H$\alpha$))
for the 244 star forming galaxies where we detect the [O\II]
emission line at a confidence $\geq 2\sigma$.

\placefigure{OII_and_Ha}

\begin{figure}
\epsscale{1.0}
\plotone{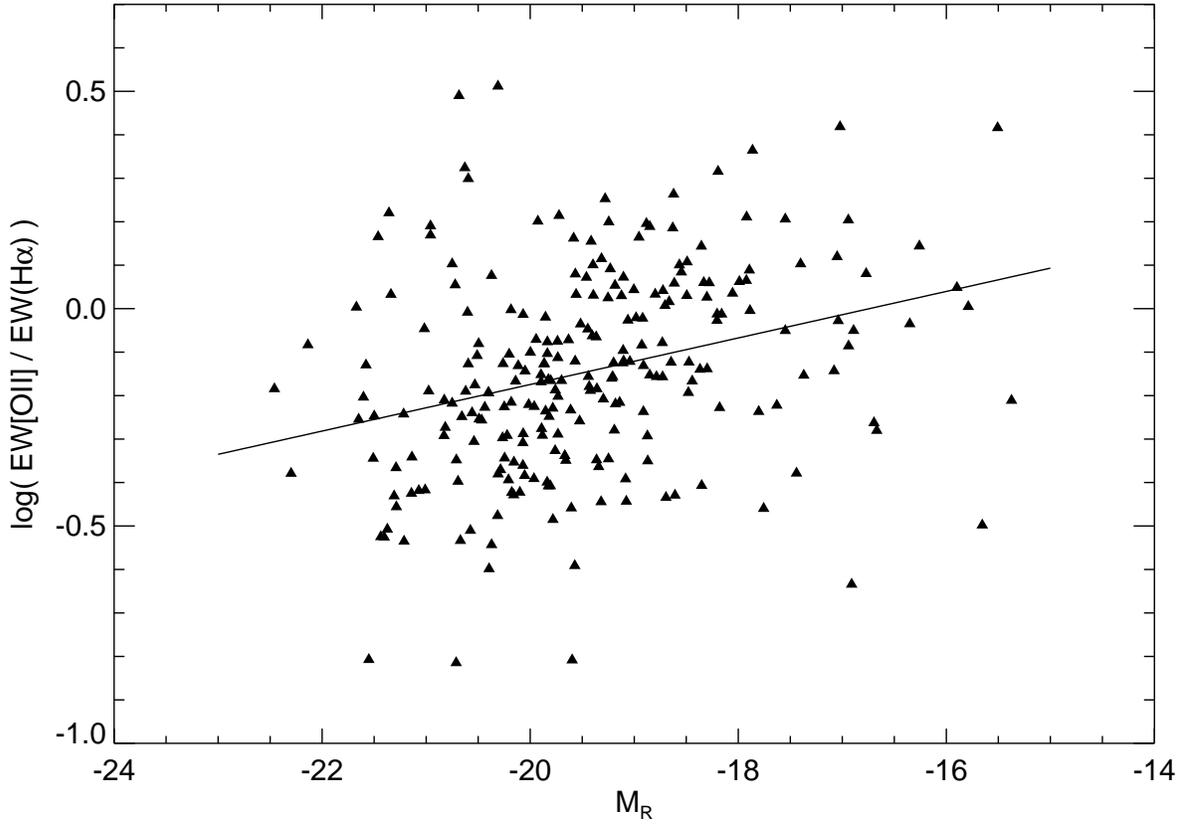}
\caption{Log of the ratio of [O\II] and H$\alpha$ EW
vs.~the $R$ absolute magnitude for the 244 galaxies in Figure
\ref{OII_and_Ha}.  The
slope of the relation (0.0054 dex mag$^{-1}$) is similar to the slope
derived by \cite{Jansen01} for the same ratio vs.~the $B$ absolute
magnitude. 
\label{OII_and_Ha_two}}
\end{figure}

Recently, \cite{Jansen01} showed that the [O\II]/H$\alpha$ ratio
correlates with luminosity. This correlation arises from reddening as
well as from the metallicity dependent excitation of the interstellar
medium.  They find a slope of 0.043 dex mag$^{-1}$ for the relation
log(EW [O\II]/H$\alpha$) vs.~M$_B$.  Our data yield a very similar
slope of 0.054 dex mag$^{-1}$ for the same relation vs.~M$_R$.  A
Spearman rank test shows that the correlation in our sample is highly
significant; the null hypothesis (no correlation) has a probability of
only $\sim$10$^{-8}$.  Figure \ref{OII_and_Ha_two} plots our log(EW
[O\II]/H$\alpha$) vs.~M$_R$ data along with the best fit linear
relation.  We confirm that the EW([O\II]) should be used with caution
as a measure of star formation \citep{Jansen01}.

\placefigure{OII_and_Ha_two}

\section{The Spectral Type -- Density Relation}

\subsection{The Density Estimator}

In analogy with the morphology-density relation \citep{Dressler,
PostmanGeller} we measure the spectral-type density relation.  The
method outlined here for measuring the density surrounding each galaxy
is similar to the one used by \cite{LCRS} and \cite{PostmanGeller}.
We determine the density from the $j$ nearest neighbors of a galaxy in
redshift space (including the galaxy in question):
$$\rho_j \equiv \frac{N_j}{V_j}$$ 
where $N_j$ is the estimated number of galaxies contained in the
volume $V_j$, with $j=10$.  We scale the number of galaxies in the
volume by a weighting function $w_i$ which accounts for the galaxies absent
from the magnitude-limited survey:
$$ N_j = \sum_{i=1}^j w_i $$
$$ w_i \equiv \frac{\int^{\infty}_{-\infty} \phi(M)dM}
{\int_{-\infty}^{M_i} \phi(M)dM} $$ 
where $M_i$ is the faintest absolute magnitude detectable at object
$i$ (eq.~[\ref{Mag}]).  Because we calibrate the 15R Survey photometry 
with the  Century Survey (CS), we use the CS
luminosity function $\phi$, with Schechter function parameters
$\phi_\ast = 0.025$ Mpc$^{-3}$, $\alpha = -1.17$, and $M_\ast =
-20.73$ \citep{Century}.

For the 15R galaxies, the weighting function is $w_i \sim 4$ at
$z_i=0.025$ and $w_i \sim 100$ at $z_i=0.075$.  We impose an upper
survey limit of $z=0.075$ to avoid very large weights and a
lower limit of $z=0.0033$.  The median radius of the regions used to
calculate densities is $\sim$5 Mpc, with a total range of 1---20 Mpc.

We use the Euclidean law of cosines to estimate the distance between
two galaxies.  Any errors introduced from this simplified relation are
small compared with the uncertainty in galaxy positions due to unknown
peculiar velocities. Because the virialized central regions of systems
of galaxies are extended along the line-of-sight in redshift space,
our prescription underestimates the density in these regions. At
intermediate densities, our prescription may overestimate the density
somewhat because of infall.  We estimate $d_{ij}$, the distance
between galaxies $i$ and $j$, from the observed angular separation
$\theta_{ij}$ and the luminosity distances to galaxies $i$ and $j$:
$$d_{ij} = 
     \left[ D_i^2 + D_j^2 - 2 D_i D_j cos(\theta_{ij}) \right]^\frac{1}{2}.$$
Typical separations between adjacent neighbors are a few Mpc.

If galaxy $k$ is sufficiently far from the survey boundaries, the
volume $V_j$ containing its $j$ nearest neighbors is simply a sphere
centered at $k$ with radius $d_{kj}$, the distance from galaxy $k$ to
its $j$-th nearest neighbor in the survey.  In general, however, the
volume $V_j$ is the intersection of this sphere with the volume
contained by the boundaries of the 15R Survey.  We therefore determine
this volume $V_j$ numerically for each galaxy.  Again, we simplify the
calculations by assuming flat Euclidean space within the volume $V_j$.

The uncertainty in $\log(\rho) \approx 0.1$.  Uncertainty in the
average number density of galaxies in the local universe contributes
some constant offset to the value of $\log(\rho)$; thus, care must be
taken when comparing number density measurements between different
surveys.  We neglect the overall survey completeness because it is, in
effect, another normalization contributing a constant offset to
$\log(\rho)$.  The small variations in survey completeness in
different regions introduce errors less than the uncertainty in
$\log(\rho)$.

Our density estimation technique avoids a number of pitfalls.  A
``smoothed'' density estimator requires careful management of the
sample boundaries, and imposing an ``average'' or ``grey'' density for
the region outside the boundaries.  Measuring the number of galaxies
within a fixed volume around each galaxy would underestimate the
density of compact clusters and overestimate the density in a void.
Scaling the volume by the local density is effectively equivalent to
the technique we adopt.

\subsection{Spectral Types and Density}
\label{st_density}

\begin{figure}
\epsscale{1.0}
\plotone{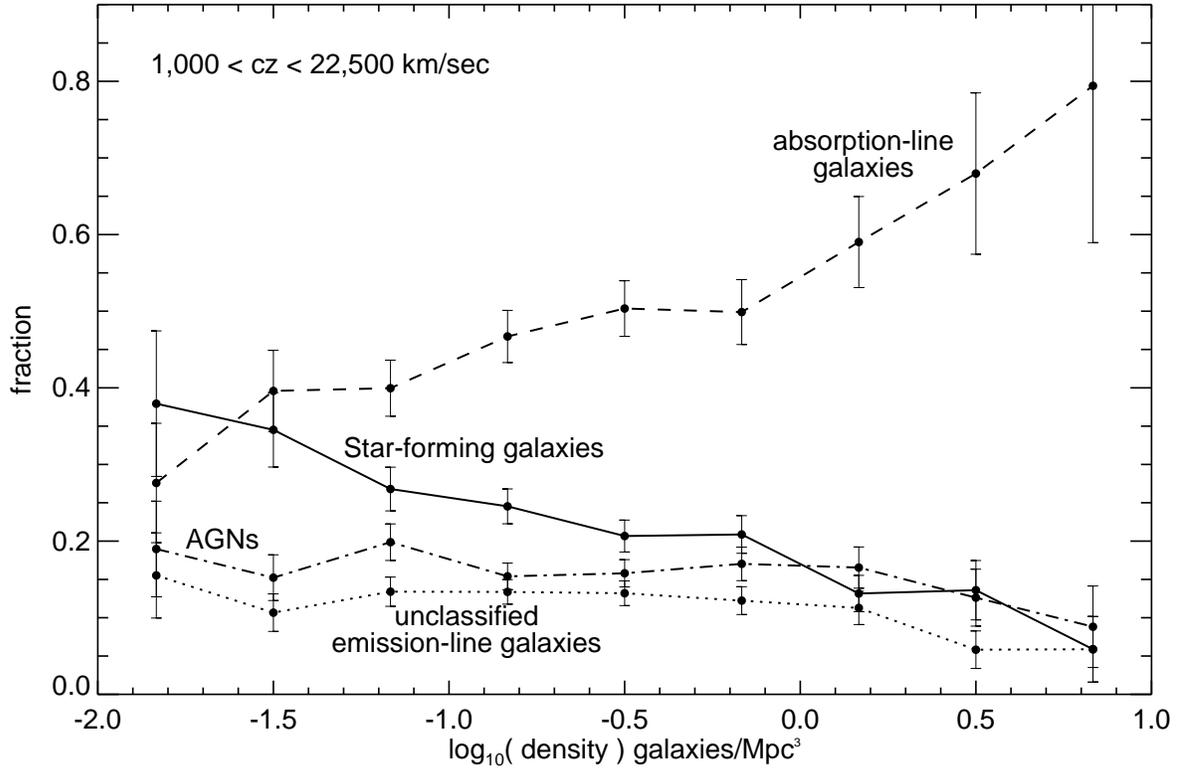}
\caption{Spectral type fractions as a function of
density.  Galaxies with star-forming spectra dominate at the lowest
densities; while galaxies with absorption line spectra dominate at
intermediate and high densities.  The fraction of unclassified
emission galaxies and AGN do not vary strongly with density.  The
first and last density bins include any galaxies outside the plot
limits.  The uncertainty in density is less than half the width of the
$\log(\rho)$ bins. We include only galaxies with 0.0033$<z<$0.075 (see
text).
\label{density}}
\end{figure}

Figure \ref{density} shows that, as expected, star-forming (H\II-like)
galaxies dominate at the lowest densities; absorption line galaxies
dominate at intermediate and high densities.  A $\chi^2$ test rejects
the hypothesis of a fraction independent of density at 99.9991\%
confidence for star-forming galaxies and 99.9914\% confidence for
absorption line galaxies.  The fractions of unclassified emission-line
galaxies and AGNs do not vary dramatically with density.  A $\chi^2$
test shows that the hypothesis of a constant fraction with density is
acceptable for AGNs, and can be rejected at 99.6\% confidence
($\sim$2.9$\sigma$) for the unclassified emission line galaxies.

Based on a smaller sample of 106 AGNs with $cz \leq 3000$ km s$^{-1}$,
\cite{Monaco} assessed the large-scale distribution of these objects.
The region they explore is dominated by the Local Supercluster making
comparison with our larger survey difficult. A qualitatively similar
conclusion of their survey and ours is that, unlike star-forming
galaxies, AGNs do occur frequently in dense regions of the
universe. The physical processes which suppress star formation in
these regions are thus ineffective in suppressing AGNs. Our larger,
deeper survey enables exploration of the AGN fraction over the full
range of galaxy densities.

\placefigure{density}

\subsection{Star Formation and Density}
\label{sf_density}

\begin{figure}
\epsscale{1.0}
\plotone{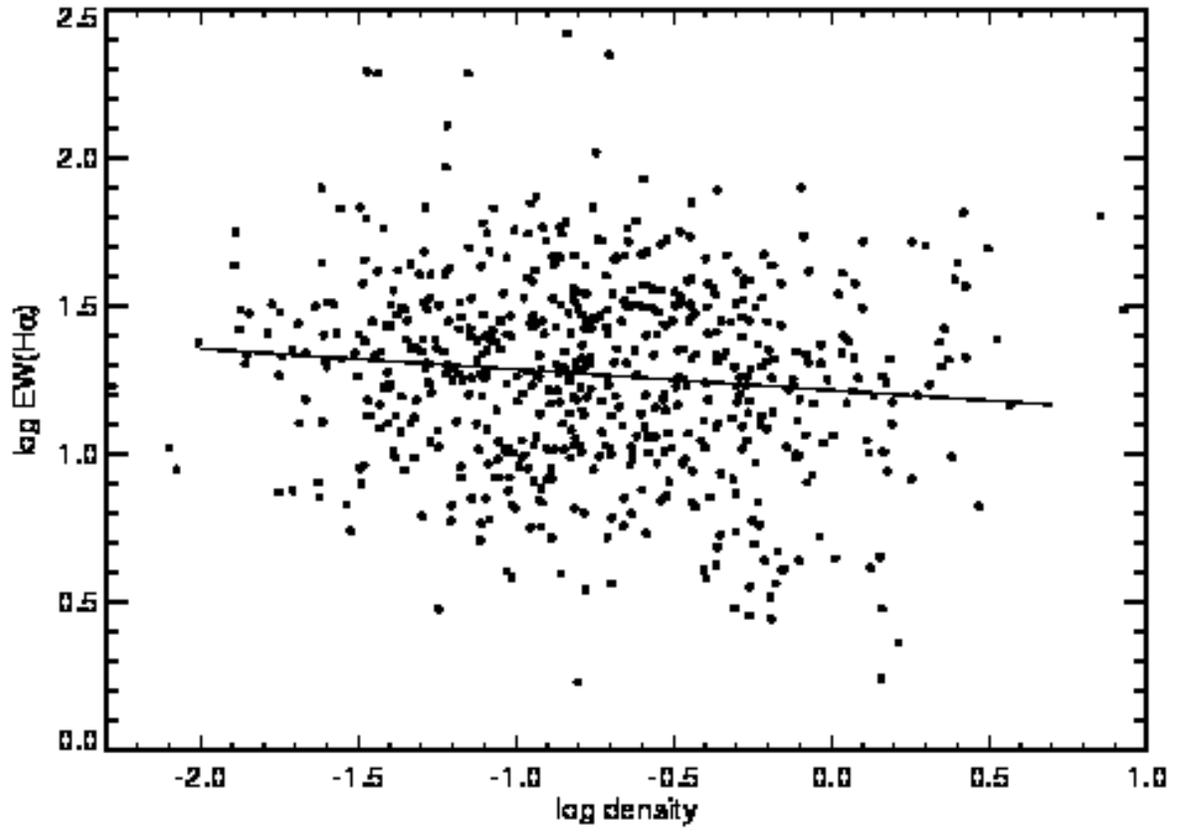}
\caption{Equivalent width of H$\alpha$ vs.~density for
the 614 star-forming galaxies with 0.0033$<z<$0.075.  The hypothesis
of no correlation can be rejected at only 98.4\% confidence
($\sim$2.4$\sigma$).
\label{HaEW_den}}
\end{figure}

We now examine whether the properties of the 641 star forming galaxies
depend on environment. Specifically we ask whether the Scalo birthrate
parameter, $b$, depends on density.  For consistency with other
studies, we use EW(H$\alpha$) instead of EW(H$\alpha$ + [N\II]); our
data show that the two are tightly correlated.  Figure \ref{HaEW_den}
shows the distribution of EW(H$\alpha$) as a function of galaxy
density for the star forming galaxies.  There is only a very
weak correlation in this plot: at high densities the EW(H$\alpha$) 
is smaller.  
The Spearman
rank test shows that the null hypothesis of no correlation can be
rejected at only 98.4\% confidence ($\sim$2.4$\sigma$).
The scatter, which is significantly larger than the slight trend, 
is nearly constant at all densities.

\placefigure{HaEW_den}
 
The weak correlation and roughly constant scatter
of EW(H$\alpha$) with densities averaged over a
scale of $\sim$5 Mpc may be related to the dependence of EW(H$\alpha$)
on spatial and velocity separation in close pairs.  \cite{BGK} argue
that the observed dependence is a consequence of tidally-triggered
star formation.  They show that in a large sample of close pairs
objectively selected from a redshift survey, the EW(H$\alpha$) declines
with projected separation and with line-of-sight velocity
difference. The decline is very steep; the maximum EW(H$\alpha$) drops
from more than 150 \AA\ at a projected separation of 5 kpc to 50 \AA\
at 40 kpc. Extrapolation of this result suggests that at the large
smoothing scales we explore here, there should be little dependence of
the EW(H$\alpha$) on density.

The correlation of EW(H$\alpha$) with density we do observe is in the
same sense but much weaker than the correlation of EW([O\II]) with
density in the Las Campanas Redshift Survey. \cite{LCRS} argue that
the ratio of the numbers of galaxies with EW([O\II]) $\ge$ 20 {\rm \AA}
to those with 5{\rm \AA} $\le$ EW([O\II]) $<$ 20{\rm \AA} declines by a
factor of two as the density increases by three orders of magnitude.
Three factors may contribute to the different LCRS result: (1) less
uniform selection in the LCRS excluding low surface brightness
galaxies which often have strong emission lines, (2) the problems of
[O\II] as a star formation indicator ($\S$3.4), and (3) aperture bias
introduced by the 3.5$^{\prime\prime}$ diameter LCRS fiber
\citep{Kochanek}. The aperture bias steepens the correlation between
EW(H$\alpha$) and surrounding galaxy density through the morphology
density relation. In denser regions, the median bulge-to-disc ratio of
galaxies is larger. Thus, with a small aperture, underestimation of
the EW(H$\alpha$) is a progressively more serious bias in denser
regions. This bias results in an apparently larger change in median
EW(H$\alpha$) with density than is actually present.

\cite{Loveday99} use the Stromlo-APM survey to examine the
relationship between EW(H$\alpha$) and galaxy density using
correlation function techniques. In contrast with the LCRS, the
Stromlo-APM data are long-slit spectra. Their typical slit covering
fraction is $\sim$0.45, about twice our typical coverage.
\cite{Loveday99} study the redshift-space autocorrelation functions of
galaxies in three bins of EW(H$\alpha$): (1) $<$2{\rm \AA}, (2)
2---15{\rm \AA}, and (3) $>$15{\rm \AA}.  Galaxies with the smallest
EW(H$\alpha$) have a significantly larger correlation length,
8.7$\pm$0.5 Mpc, than the galaxies with stronger emission, which have
correlation lengths of 5.5$\pm$0.7 Mpc and 4.6$\pm$0.9 Mpc,
respectively.  We classify most galaxies with EW(H$\alpha$) $<$2{\rm
\AA} as absorption line galaxies, and we have shown in $\S$4.2 that
these dominate at the highest densities.  \cite{Loveday99} find no
significant difference in the redshift-space autocorrelation length
for their two groups of emission line galaxies, in qualitative
agreement with our Figure \ref{HaEW_den}. We conclude that for
emission-line galaxies, the dependence of EW(H$\alpha$) on local
galaxy density in redshift space is weak and the the strong dependence
claimed in the LCRS is a probable result of aperture bias.

\section{Conclusion}

15R-North is a red-selected, magnitude-limited sample of galaxies at
low redshift ($z \le 0.15$).  With uniform optical spectra for 90\% of
the 15R-North galaxies, we have a complete spectroscopic census of the
local universe: 51\% of the galaxies have absorption-line spectra and
49\% have emission.  At least 20\% are H\II-like and 17\% have
AGN-like spectra.  If all of the unclassified emission-line spectra
(the remaining 12\% of the sample) were AGN spectra, AGNs would make
up 28\% of the galaxies in our sample, probably consistent with the
43\% identified by \cite{Ho97} in their blue-selected survey. Because
\cite{Ho97} base their conclusions on high S/N nuclear spectra and
ours typically cover a much larger fraction of the galaxy (diluting
the AGN contribution), \cite{Ho97} should estimate a larger AGN
fraction than we do.  The consistently large AGN fraction may
complicate evaluation of the star formation rate at zero redshift.

AGN and star-formation activity have a different dependence on the
surrounding galaxy number density (averaged over a several Mpc scale).
The AGN fraction is independent of local density; the star-forming
fraction decreases steeply as the galaxy density increases.  The
unclassifiable galaxies have a negligible effect on these results;
their fraction decreases slowly as the density increases. AGNs are
more abundant among the more luminous galaxies in the survey;
star-forming galaxies are more abundant in low-luminosity
systems. These trends may be useful in unraveling the AGN-starburst
connection.

Perhaps the most surprising result of analysis of the 15R data is the
very shallow dependence of the EW(H$\alpha$) in star-forming galaxies
on the surrounding density in redshift space. This result is actually
similar to the conclusions drawn by \cite{Loveday99} from their
longslit spectroscopy in the Stromlo-APM survey. We suggest that this
shallow dependence implies that processes occurring on much smaller
scales (e.g.\ tidal triggering) determine the EW(H$\alpha$).

Examination of the relative distributions of absorption- and
emission-line galaxies recovers the result obtained by others:
absorption-line galaxies are more abundant in dense regions and
emission-line galaxies are more abundant at low densities. Thus the
gross distinction in spectroscopic properties is sensitive to the
larger environment. This relation is, at least in part, a consequence
of the well-known morphology-density relation.  Elliptical and S0
galaxies, which usually have absorption-line spectra, inhabit the
densest regions of the universe.
 
\acknowledgments
G. Bothun made extensive comments on an early draft of this paper, and
we thank him for his suggestions which enormously strengthened the paper.
We thank N. Kashikawa for allowing us to use his Coma Cluster
photometry in advance of publication.  We also thank R. Kron for
several valuable suggestions.  We are grateful to P. Berlind for
making many of the FAST observations and S. Tokarz for the pipeline
reductions.  B. C. was supported by a Predoctoral Fellowship from the
Smithsonian Astrophysical Observatory.

The POSS data were based on photographic data of the National
Geographic Society---Palomar Geographic Society to the California
Institute of Technology.  The plates were processed into the present
compressed digital form with their permission. The Digitized Sky
Survey was produced at the Space Telescope Science Institute under US
Government grant NAG W-2166.

\clearpage

\clearpage

\clearpage

\clearpage

\clearpage

\clearpage

\clearpage

\clearpage

\clearpage

\clearpage

\clearpage

\clearpage

\clearpage
 
\clearpage

\clearpage

\clearpage

\clearpage

\clearpage

\clearpage

\clearpage
\begin{figure}
\epsscale{.7}
\plotone{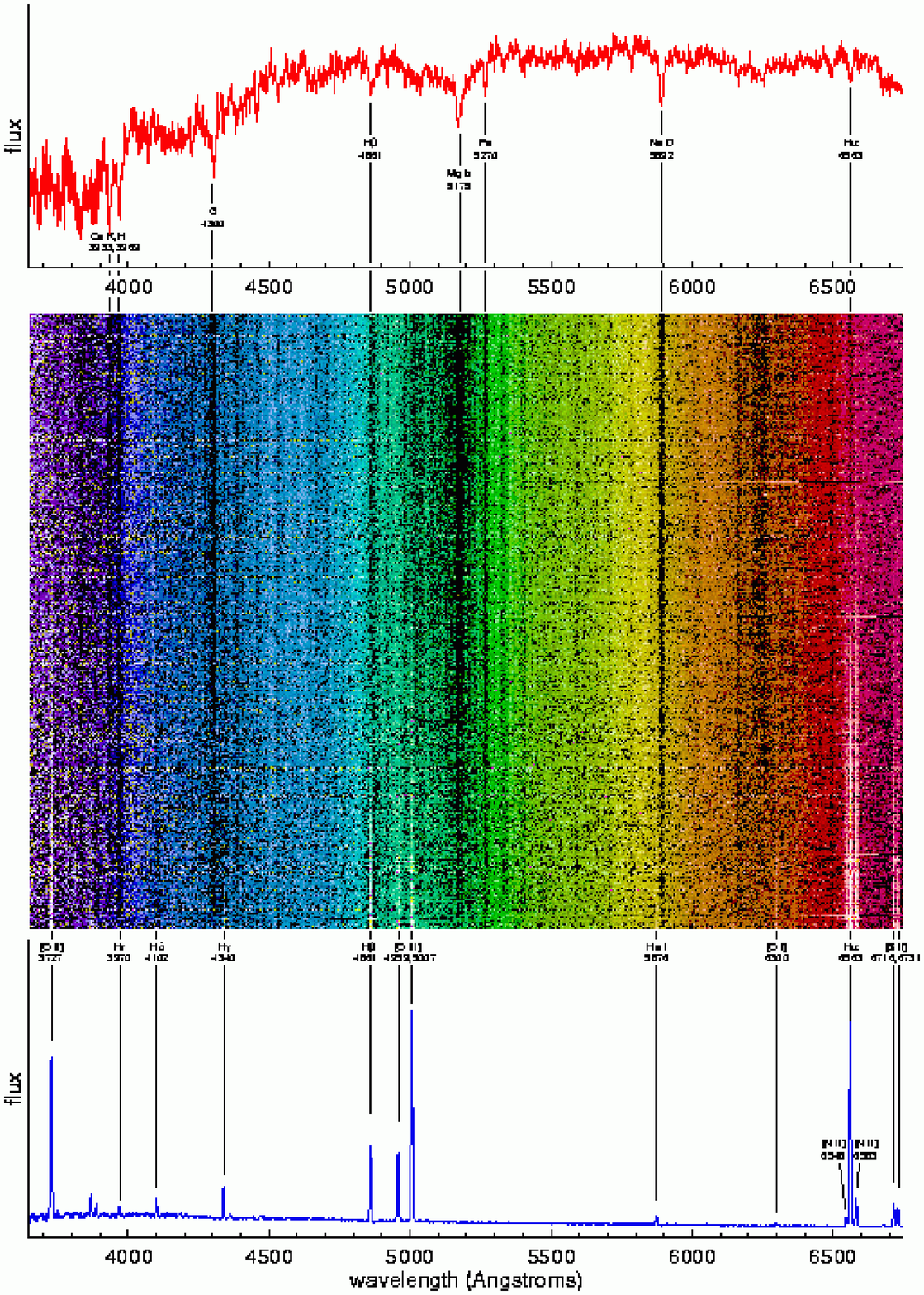}
\caption{Plate 1.---The 15R spectra.  The abscissa is rest frame wavelength.  
We display all the spectra in order of EW(H$\alpha$), 
with largest EW(H$\alpha$) at the bottom.
We remove the continuum from each spectrum to emphasize the spectral features.
Stellar absorption features (dark vertical bands) are common to all the 
spectra;  emission features (pale vertical bands at the bottom of
the image) appear in about half of the galaxies.
The example spectra at the top and bottom are averages of 
10 of the strongest absorption and emission spectra, respectively,
without continuum removal.}
\end{figure}

\clearpage
\begin{table} \caption{Line Index Definitions\label{lines}}
\begin{tabular}{rlccc}\tableline\tableline
\multicolumn{2}{c}{Index} & Blue Continuum & Line Region & Red Continuum \\
\ & \  & (\AA) & (\AA) & (\AA) \\ \tableline
{[}O\II{]}  & 3727.3\AA & 3653.0---3716.3 & 3716.3---3738.3 & 3738.3---3803.0 \\
H$\beta$  & 4861.3\AA & 4761.3---4841.3 & 4841.3---4881.3 & 4881.3---4961.3 \\
{[}O\III{]} & 5006.8\AA & 4891---4945     & 4995---5019     & 5021---5087 \\
{[}N\II{]}  & 6548.1\AA & 6505---6535     & 6538.1---6558.1 & 6597---6627 \\
H$\alpha$ & 6562.8\AA & 6505---6535     & 6554.5---6574.5 & 6597---6627 \\
{[}N\II{]}  & 6583.4\AA & 6505---6535     & 6573.4---6593.4 & 6597---6627 \\ \tableline
\end{tabular}
\end{table}

\clearpage
\begin{table} \caption{Spectral Types in the 15R Sample\label{types}}
\begin{tabular}{rl}\tableline\tableline
Number & Type \\ \tableline
641 & H\II-type emission \\
522 & AGN-type emission \\
375 & unclassified emission \\
1611 & absorption-line spectra \\ \tableline
3149 & total galaxies \\ \tableline
\end{tabular}
\end{table}

\clearpage
\begin{table} \caption{The Most Vigorous Star-Forming Galaxies in 
the 15R Sample\label{vigorous-sf}}
\begin{tabular}{rlrrr}\tableline\tableline
RA (1950) & Dec (1950) & \  & \multicolumn{1}{c}{$cz$} & EW(H$\alpha$)\\
(hh:mm:ss) & (deg:mm:ss) & \multicolumn{1}{c}{$b$} & \multicolumn{1}{c}{(km/s)} & \multicolumn{1}{c}{(\AA)}\\ \tableline
  13:28:21.4 &  31:32:27.2 N  & 18.8 &  10198 &   262.5 \\
  11:54:54.0 &  31:21:43.6    & 11.2 &   6916 &   223.7 \\
  14:52:07.7 &  30:24:09.4    & 10.1 &   2859 &   196.4 \\
  14:52:04.8 &  30:24:40.3     & 9.6 &   2833 &   193.2 \\
  14:55:28.6 &  26:51:51.8     & 8.7 &   1267 &   193.1 \\
   9:47:26.4 &  31:43:18.1     & 7.4 &    524 &   167.5 \\
  12:35:12.8 &  27:24:13.3     & 6.7 &   4593 &   128.8 \\
  16:33:43.2 &  27:35:17.5     & 5.1 &  14881 &    93.4 \\
  12:42:49.7 &  27:23:54.6     & 4.9 &   1094 &   104.2 \\
  10:43:04.8 &  27:52:58.4     & 4.6 &  13311 &    68.2 \\
  14:22:07.3 &  31:38:24.7 W   & 4.5 &  12026 &    79.1 \\
  12:44:28.3 &  26:50:13.6     & 4.4 &    800 &    92.1 \\
  12:13:23.5 &  26:56:23.3     & 4.2 &   7627 &    85.0 \\
   8:59:42.3 &  31:28:18.1     & 4.1 &   4141 &    67.3 \\
  13:24:29.9 &  26:51:01.1 W   & 4.1 &   7013 &    70.6 \\
  13:35:57.5 &  28:01:30.7     & 3.7 &   9996 &    67.6 \\
  13:17:27.6 &  31:05:15.4     & 3.7 &   7299 &    79.3 \\
  13:28:21.4 &  31:32:27.2 S   & 3.7 &  10187 &    58.6 \\
  16:33:43.2 &  27:35:17.5 E   & 3.7 &  14742 &    68.2 \\
   8:32:28.4 &  30:42:27.0     & 3.5 &   7693 &    68.3 \\
  11:45:13.3 &  31:37:21.7     & 3.4 &   8356 &    70.2 \\
  11:20:26.9 &  30:45:11.2     & 3.4 &   1664 &    73.9 \\
   9:39:56.9 &  32:04:34.0     & 3.4 &   1331 &    57.0 \\
  14:35:42.4 &  30:41:42.4     & 3.4 &  10396 &    56.0 \\
  13:13:28.2 &  26:49:10.2     & 3.3 &  11484 &    58.2 \\
  11:06:56.2 &  27:11:31.6     & 3.3 &  21091 &    52.4 \\
  12:17:05.3 &  32:28:40.8     & 3.3 &  31400 &    61.2 \\
  12:56:10.3 &  27:32:03.1     & 3.3 &   7344 &    65.5 \\
  11:02:55.3 &  31:06:09.0     & 3.2 &  10205 &    54.2 \\
   9:02:01.2 &  28:32:48.8     & 3.1 &  14564 &    56.4 \\
  14:46:00.8 &  27:34:41.2     & 3.1 &  15440 &    53.2 \\
   9:47:17.9 &  28:14:49.2     & 3.1 &   1450 &    77.9 \\
   9:27:31.3 &  27:59:44.5     & 3.0 &  12992 &    54.4 \\ \tableline
\end{tabular}
\end{table}

\end{document}